\documentclass[11pt]{article} 
\usepackage{latexsym}
\usepackage{lscape}
\usepackage{url}
\usepackage{amssymb}
\usepackage{verbatim}
\usepackage[pdftex]{graphicx}   

\normalsize
\makeatletter
\@addtoreset{equation}{section}
\makeatother

\newcommand{\myuparrow}{\hspace*{-1mm}\uparrow}

\newcommand{\wrt}{w.r.t.}

\newtheorem{theorem}{Theorem}[section]
\newtheorem{lemma}[theorem]{Lemma}

\newenvironment{proof}{{\sc Proof. }}{\hfill$\Box$\vspace{0.1in}}
\title{A Cubic-Time 2-Approximation Algorithm for rSPR Distance}

\author{
Zhi-Zhong Chen\thanks{
Corresponding author. 
Division of Information System Design, Tokyo Denki University, 
Ishizaka, Hatoyama, Hiki, Saitama, 359-0394, Japan.
Email: {\tt zzchen@mail.dendai.ac.jp}. 
Phone: +81-49-296-5249. Fax: +81-49-296-7072.
}
\and
Eita Machida\thanks{Division of Information System Design, 
Tokyo Denki University, Ishizaka, Hatoyama, Hiki, Saitama, 359-0394, Japan.
Email: {\tt 16rmd27@ms.dendai.ac.jp}. 
}
\and
Lusheng Wang\thanks{Department of Computer Science,
City University of Hong Kong, 
Tat Chee Avenue, Kowloon, Hong Kong SAR.
Email: {\tt cswangl@cityu.edu.hk}.
Phone: +852-2788-9820. Fax: +852-2788-8614.
    }
    }

\date{}

\begin{document}

\maketitle{}
\begin{abstract}
Due to hybridization events in evolution, studying two different genes of 
a set of species may yield two related but different phylogenetic trees for 
the set of species. In this case, we want to measure the dissimilarity of 
the two trees. The rooted subtree prune and regraft (rSPR) distance of 
the two trees has been used for this purpose. The problem of computing the 
rSPR distance of two given trees has many applications but is 
NP-hard. The previously best approximation algorithm for rSPR distance achieves 
a ratio of 2 in {\em polynomial} time and its analysis is based on the duality theory 
of linear programming. In this paper, we present a {\em cubic}-time approximation 
algorithm for rSPR distance that achieves a ratio of $2$. Our algorithm is based on 
the notion of {\em key} and several structural lemmas; its analysis is purely 
combinatorial and explicitly uses a search tree for computing rSPR distance exactly. 
\end{abstract}


\section{Introduction}\label{sec:intro}
When studying the evolutionary history of a set $X$ of existing species, one can obtain 
a phylogenetic tree $T_1$ with leaf set $X$ with high confidence by looking at a segment 
of sequences or a set of genes~\cite{MWZ99,MZ11}. When looking at another segment of 
sequences, a different phylogenetic tree $T_2$ with leaf set $X$ can be obtained  with 
high confidence, too. In this case, we want to measure the dissimilarity of $T_1$ and $T_2$. 
The rooted subtree prune and regraft (rSPR) distance between $T_1$ and $T_2$ has been used 
for this purpose \cite{h+1996}. It can be defined as the minimum number of edges that should 
be deleted from each of $T_1$ and $T_2$ in order to transform them into {\em essentially 
identical} rooted forests $F_1$ and $F_2$. Roughly speaking, $F_1$ and $F_2$ are {\em 
essentially identical} if they become identical forests (called {\em agreement forests} of 
$T_1$ and $T_2$) after repeatedly contracting an edge $(p,c)$ in each of them such that 
$c$ is the unique child of $p$ (until no such edge exists). 

The rSPR distance is an important metric that often helps us discover reticulation events. 
In particular, it provides a lower bound on the number of reticulation events \cite{Bar+05,BH06}, 
and has been regularly used to model reticulate evolution \cite{Mad97,Nak+05}. 

Unfortunately, it is NP-hard to compute the rSPR distance of two given phylogenetic 
trees~\cite{BS05,h+1996}. This has motivated researchers to design approximation 
algorithms for the problem~\cite{B+2006,BMS08,h+1996,RSW07}. Hein {\em et al.} \cite{h+1996} 
were the first to come up with an approximation algorithm. They also introduced the important 
notion of maximum agreement forest (MAF) of two phylogenetic trees. Their algorithm was 
correctly analyzed by Bonet {\em et al.} \cite{B+2006}. Rodrigues {\em et al.} \cite{RSW07} 
modified Hein {\em et al.}'s algorithm so that it achieves an approximation ratio of~3 and 
runs in quadratic time. Whidden {\em et al.} \cite{WZ09} came up with a very simple 
approximation algorithm that runs in linear time and achieves an approximation ratio of~3. 
Although the ratio~3 is achieved by a very simple algorithm in \cite{WZ09}, 
no polynomial-time approximation algorithm had been designed to achieve a better ratio 
than~3 before Shi {\em et al.} \cite{Shi+14} presented a polynomial-time approximation 
algorithm that achieves a ratio of 2.5. Schalekamp {\em et al.} \cite{Svv16} presented 
a {\em polynomial}-time 2-approximation algorithm for the same problem. However, they use 
an LP-model of the problem and apply the duality theory of linear programming in the analysis 
of their algorithm. Hence, their analysis is not intuitively understandable. Moreover, 
they did not give an explicit upper bound on the running time of their algorithm. 
Unaware of Schalekamp {\em et al.}'s work \cite{Svv16}, we \cite{CMW16} presented 
a {\em quadratic}-time $\frac{7}{3}$-approximation algorithm for the problem; 
the algorithm is relatively simpler and its analysis is purely combinatorial. 

In certain real applications, the rSPR distance between two given phylogenetic trees 
is small enough to be computed exactly within reasonable amount of time. 
This has motivated researchers to take the rSPR distance as a parameter and 
design fixed-parameter algorithms for computing the rSPR distance of 
two given phylogenetic trees~\cite{BS05,CFW15,Wu09,WZ09,WBZ10}. These algorithms are 
basically based on the branch-and-bound approach and use the output of an approximation 
algorithm (for rSPR distance) to decide if a branch of the search tree should be cut. Thus, 
better approximation algorithms for rSPR distance also lead to faster exact algorithms for 
rSPR distance. It is worth noting that approximation algorithms for rSPR distance can also 
be used to speed up the computation of hybridization number and the construction of minimum 
hybridization networks~\cite{CW12,CW13}. 

In this paper, we improve our $\frac{7}{3}$-approximation algorithm in \cite{CMW16} 
to a new 2-approximation algorithm. Our algorithm proceeds in stages until the input 
trees $T_1$ and $T_2$ become identical forests. Roughly speaking, in each stage, 
our algorithm carefully chooses a dangling subforest $S$ of $T_1$ and uses $S$ to 
carefully choose and remove a set $B$ of edges from $T_2$. $B$ has a crucial property 
that the removal of the edges of $B$ decreases the rSPR distance of $T_1$ and $T_2$ by 
at least~$\frac{1}{2}|B|$. Because of this property, our algorithm achieves a ratio of $2$.
As in \cite{CMW16}, the search of $S$ and $B$ in our algorithm is based on our original 
notion of {\em key}. However, unlike the algorithm in \cite{CMW16}, the subforest $S$ 
in our new algorithm is not bounded from above by a constant. This difference is crucial, 
because the small bounded size of $S$ in \cite{CMW16} makes for a tedious case-analysis. 
Fortunately, we can prove a number of structural lemmas which enable us to 
construct $B$ systematically and hence avoid complicated case-analysis. Our analysis of 
the algorithm explicitly uses a search tree (for computing the rSPR distance of two given 
trees exactly) as a tool, in order to show that it achieves a ratio of~2. 
To our knowledge, we were the first to use a search tree explicitly for this purpose. 

The remainder of this paper is organized as follows. 
Section~\ref{sec:prob} reviews the rSPR distance problem and presents our main theorems. 
Section~\ref{sec:def} gives the basic definitions that will be used thereafter. 
Section~\ref{sec:search} shows how to build a search tree for computing the rSPR distance exactly. 
Section~\ref{sec:key} defines the important notion of {\em key}. Section~\ref{sec:goodKey} 
presents our algorithm for finding a good key or cut.

\section{The rSPR Distance Problem and the Main Theorems}\label{sec:prob}
In this paper, a forest $F$ always means a rooted forest in which each vertex has at most 
two children. A vertex of $F$ is {\em unifurcate} (respectively, {\em bifurcate}) if its number 
of children in $F$ is 1 (respectively, 2). 
If a vertex $v$ of $F$ is not a root in $F$, then $e_F(v)$ denotes the edge entering $v$ in $F$;
otherwise, $e_F(v)$ is undefined. 
%
For a set or sequence $U$ of vertices, $\ell_{F}(U)$ denotes the lowest common ancestor (LCA) of 
the vertices of $U$ in $F$ if the vertices of $U$  are in the same connected component of $F$, 
while $\ell_F(U)$ is undefined otherwise. 
When $\ell_F(U)$ is defined, we simply use $e_F(U)$ to denote $e_F(\ell_F(U))$. 

$F$ is {\em binary} if it has no unifurcate vertex. {\em Binarizing} $F$ is 
the operation of modifying $F$ by repeatedly contracting an edge between a unifurcate vertex $p$ 
and its unique child $c$ into a single vertex $c$ until no vertex of $F$ is unifurcate. 
The {\em binarization} of $F$ is the forest obtained by binarizing $F$. 

\begin{figure}[htb]
\centerline{\includegraphics{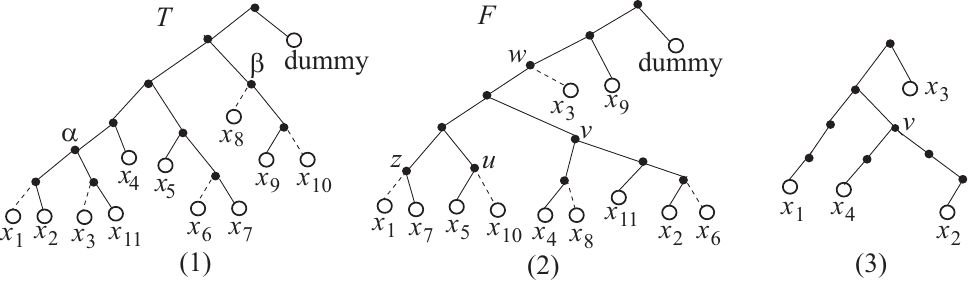}}
\caption{(1)~A phylogeny $T$, \ (2)~another phylogeny $F$, \ (3)~$F\myuparrow_{\{x_1,\ldots,x_4\}}$. 
}
\label{fig:def}
\end{figure}

A {\em phylogenetic forest} is a binary forest $F$ whose leaves are distinctively labeled but 
non-leaves are unlabeled. Whenever we say that two phylogenetic forests are isomorphic, we always 
mean that the bijection also respects the leaf labeling. $F$ is a {\em phylogeny} if it is connected. 
Figure~\ref{fig:def} shows two phylogenies $T$ and $F$. For convenience, we allow {\em the empty 
phylogeny} (i.e., the phylogeny without vertices at all) and denote it by $\bot$. For two 
phylogenetic forests $F_1$ and $F_2$ with the same set of leaf labels, we always view two leaves 
of $F_1$ and $F_2$ with the same label as the same vertex although they are in different forests. 

\begin{figure}[htb]
\centerline{\includegraphics{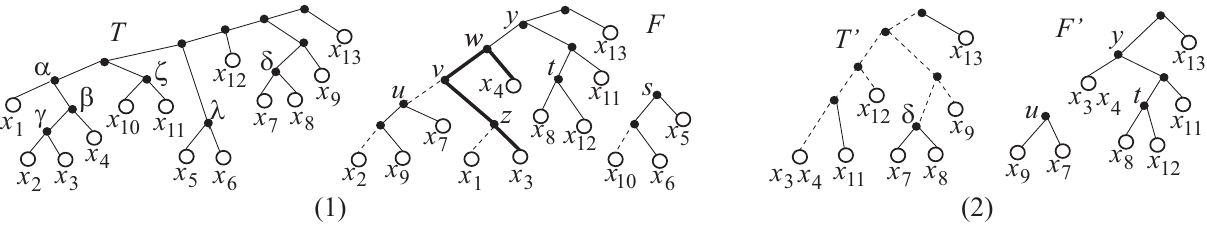}}
\caption{(1)~A TF-pair $(T,F)$ \  \ (2)~an induced sub-TF pair $(T',F')$ of $(T,F)$.}
\label{fig:conf}
\end{figure}

Suppose that $F$ is a phylogenetic forest and $C$ is a set of edges in $F$. 
$F-C$ denotes the forest obtained from $F$ by deleting the edges in $C$. 
$F-C$ may not be phylogenetic, because it may have unlabeled leaves or unifurcate vertices. 
$F \ominus C$ denotes the phylogenetic forest obtained from $F-C$ by first removing all vertices 
without labeled descendants and then binarizing it. 
$C$ is a {\em cut} of $F$ if each connected component of $F-C$ has a labeled leaf. 
If in addition every leaf of $F-C$ is labeled, then $C$ is a {\em canonical cut} of $F$. 
For example, if $F$ is as in Figure~\ref{fig:conf}(1), then the four dashed edges in $F$ 
form a canonical cut of $F$. It is known that if $C$ is a set of edges in $F$, then 
$F$ has a canonical cut $C'$ such that $F \ominus C = F \ominus C'$ \cite{BMS08}. 

Given a cut $\hat{C}$ of $F\ominus C$, we can extend $C$ to a cut $C'$ of $F$ with 
$|C'| = |C| + |\hat{C}|$ as follows. Clearly, $F-C$ is a subgraph of $F$. Let $F'$ be 
the forest obtained from $F-C$ by removing all vertices without labeled descendants. 
Obviously, $F\ominus C$ can be transformed back into $F'$ by replacing each edge $(u,v)$ 
of $F\ominus C$ with the path $P_{u,v}$ from $u$ to $v$ in $F$. 
Note that each vertex of $P_{u,v}$ other than $u$ and $v$ is unifurcate in $F'$. 
For convenience, we refer to the edge of $P_{u,v}$ leaving $u$ as {\em the edge of $F$ 
corresponding to the edge $(u,v)$ in $F\ominus C$}. One can see that $C'=C\cup \tilde{C}$ 
is a cut of $F$, where $\tilde{C}$ consists of the edges in $F$ corresponding to the edges 
in $\hat{C}$. Moreover, if both $C$ and $\hat{C}$ are canonical, then so is $C'$. 

\medskip
{\sf The rSPR Distance Problem:} Given a pair $(T,F)$ of phylogenies with the same 
set of leaf-labels, find a cut $C_T$ in $T$ and a {\em smallest} cut $C_F$ in $F$ such that 
$T\ominus C_T$ and $F\ominus C_F$ are isomorphic.
\medskip

For example, if $T$ and $F$ in Figure~\ref{fig:def} are the input to the rSPR distance problem, 
then the dashed edges in $T$ and those in $F$ together form a possible output. In the above 
definition, we require that the size of $C_F$ be minimized; indeed, it is equivalent to require 
that the size of $C_T$ be minimized because the output $C_T$ and $C_F$ have the same size. 

To solve the rSPR distance problem, it is more convenient to relax the problem by only requiring 
that $T$ be a phylogenetic tree (i.e., a connected phylogenetic forest) and $F$ be a phylogenetic 
forest. Hereafter, we assume that the problem has been relaxed in this way. Then, we refer to 
each input $(T,F)$ to the problem as a {\em tree-forest (TF) pair}. 
The size of the output $C_F$ is the {\em rSPR distance} of $T$ and $F$, and is denoted by $d(T,F)$. 
In the sequel, we assume that a TF-pair $(T,F)$ always satisfies that no leaf of $F$ is a root of $F$. 
This assumption does not lose generality, because we can remove $x$ from both $T$ and $F$ without 
changing $d(T,F)$ if $x$ is both a leaf and a root of $F$. By this assumption, 
$e_F(x)$ is defined for all leaves $x$ in $F$. We also assume that no two leaves $x$ and 
$y$ are siblings in both $T$ and $F$. This assumption does not lose generality, because we can 
contract the subtree of $T$ (respectively, $F$) rooted at the parent of $x$ and $y$ into a single 
leaf (whose label is, say the concatenation of those of $x$ and $y$) without changing $d(T,F)$. 

It is worth pointing out that to compute $d(T,F)$, it is required in the literature that 
we preprocess each of $T$ and $F$ by first adding a new root and a {\em dummy} leaf and further 
making the old root and the {\em dummy} be the children of the new root. However, the common 
{\em dummy} in the modified $T$ and $F$ can be viewed as an ordinary labeled leaf and hence 
we do not have to explicitly mention the {\em dummy} when describing an algorithm. 

To compute $d(T,F)$ for a given TF-pair $(T,F)$, it is unnecessary to compute both 
a cut $C_T$ in $T$ and a cut $C_F$ in $F$. Indeed, it suffices to compute only $C_F$, because 
a cut in $F$ forces a cut in $T$. To make this clear, we obtain a cut $C_T$ of $T$ from 
a (possibly empty) cut $C_F$ of $F$ as follows. 
\begin{enumerate}
\item\label{step:initC_T} 
	Initially, $C_T = \emptyset$. 
\item\label{step:force}
	While $T\ominus C_T$ has a non-root $\alpha$ such that the subtree of $T\ominus C_T$ rooted at 
	$\alpha$ is isomorphic to a connected component of $F\ominus C_F$, add to $C_T$ the edge of $T$ 
	corresponding to $(\beta,\alpha)$, where $\beta$ is the parent of $\alpha$ in $T\ominus C_T$. 
\end{enumerate}

We refer to $C_T$ obtained from $C_F$ as above as {\em the cut of $T$ forced by $C_F$}. 
For convenience, we say that a vertex $\alpha$ of $T \ominus C_T$ and a vertex $v$ of $F\ominus C_F$
{\em agree} if the subtree of $T \ominus C_T$ rooted at $\alpha$ is isomorphic to the subtree of 
$F \ominus C_F$ rooted at $v$. For example, if $C_F$ consists of the four dashed edges in 
Figure~\ref{fig:conf}(1), then $C_T = \{e_F(x_1), e_F(x_2), e_F(x_{10}), e_F(\lambda)\}$, 
$\beta$ in $T \ominus C_T$ and $w$ in $F \ominus C_F$ agree, and so do $\lambda$ in 
$T \ominus C_T$ and $s$ in $F \ominus C_F$. We further define the {\em sub-TF pair of $(T,F)$ 
induced by $C_F$} to be the TF pair $(T', F')$ obtained as follows. 
\begin{enumerate}
\item\label{step:init} 
	Initially, $F' = F \ominus C_F$ and $T' = T\ominus C_T$, where $C_T$ is the cut of $T$ 
	forced by $C_F$. 

\item\label{step:single}
	For each connected component $K$ of $F'$ isomorphic to a connected component $H$ of $T'$, 
	delete $K$ and $H$ from $F'$ and $T'$, respectively. 

\item\label{step:find}
	Find the set $A$ of pairs $(\alpha,v)$ such that $\alpha$ and $v$ are non-leaves in $T'$ and $F'$ 
	respectively and they agree but their parents do not. 

\item\label{step:clean2} 
	For each $(\alpha,v)\in A$ (in any order), modify $T'$ (respectively, $F'$) by contracting the subtree 
	rooted at $\alpha$ (respectively, $v$) into a single leaf $\tilde{\alpha}$ (respectively, $\tilde{v}$) 
	and further assigning the same new label to $\tilde{\alpha}$ and $\tilde{v}$. 
\end{enumerate}
For example, if $C_F$ consists of the four dashed edges in Figure~\ref{fig:conf}(1), then 
the sub-TF pair induced by $C_F$ is as in Figure~\ref{fig:conf}(2). We can view $T'$ 
(respectively, $F'$) as a subgraph of $T\ominus C_T$ (respectively, $F\ominus C_F$), by viewing 
$\tilde{\alpha}$ (respectively, $\tilde{v}$) as $\alpha$ (respectively, $v$) for each $(\alpha,v)\in A$. 

If $(T',F') = (\bot,\bot)$, then $C_F$ is an {\em agreement cut} of $(T,F)$ and $F\ominus C_F$ 
is an {\em agreement forest} of $(T,F)$. If in addition, $C_F$ is canonical, then $C_F$ is 
a {\em canonical agreement cut} of $(T,F)$. 
The smallest size of an agreement cut of $(T,F)$ is actually $d(T,F)$. 

To compute an approximation of $d(T,F)$, our idea is to look at a local structure 
of $T$ and $F$ and find a cut within the structure. A cut $C$ of $F$ is {\em good} 
(respectively, {\em fair}) for $(T,F)$ if $d(T', F') \le d(T,F) - \frac{1}{2}|C|$ 
(respectively, $d(T', F') \le d(T,F) - \frac{1}{2}(|C|-1)$), where $(T',F')$ is 
the sub-TF pair of $(T,F)$ induced by $C$. It is trivial to find a fair cut $C$ 
for $(T,F)$ with $|C|=3$ \cite{WZ09}. In contrast, Theorem~\ref{th:goodRatio} is hard to 
prove and the next sections are devoted to its proof. 

\begin{theorem}\label{th:goodRatio}
Given a TF-pair $(T,F)$, we can find a good cut for $(T,F)$ in quadratic time.
\end{theorem}

\begin{theorem}\label{th:main}
Given a TF-pair $(T,F)$, we can compute an agreement cut $C_F$ of $(T,F)$ with 
$|C_F| \le 2d(T,F)$ in cubic time. 
\end{theorem}
\begin{proof}
Given a TF-pair $(T,F)$, we first compute a good cut $C$ for $(T,F)$ and then construct the sub-TF 
pair $(T',F')$ of $(T,F)$ induced by $C$. If $(T',F') = (\bot, \bot)$, then we return $C$; 
otherwise, we recursively compute an agreement cut $C'$ of $(T',F')$, and then return $C \cup C''$, 
where $C''$ consists of the edges in $F$ corresponding to those in $C'$. 

The correctness and time complexity of the above algorithm are clear. We next prove that 
the algorithm returns an agreement cut $C_F$ of $(T,F)$ with $|C_F| \le 2d(T,F)$ by induction 
on the recursion depth of the algorithm. If the algorithm makes no recursive call, it is clear 
that the algorithm outputs an agreement cut $C$ of $(T,F)$ with $|C| \le 2d(T,F)$. So, assume 
that the algorithm makes a recursive call. Then, $|C'| \le 2d(T',F')$ by the inductive hypothesis. 
Now, since $d(T',F') \le d(T,F) - \frac{1}{2}|C|$, $|C\cup C''| = |C| + |C'| \le 
2(d(T,F) - d(T',F')) + 2d(T',F') = 2d(T,F)$.
\end{proof}

\section{Basic Definitions and Notations}\label{sec:def}
Throughout this section, let $F$ be a phylogenetic forest. 
We view each vertex $v$ of $F$ as an ancestor and descendant of itself. 
Two vertices $u$ and $v$ of $F$ are {\em comparable} if $\ell_F(u,v) \in \{u,v\}$, 
while they are {\em incomparable} otherwise. 
For brevity, we refer to a connected component of $F$ simply as a {\em component} of $F$. 
We use $L(F)$ to denote the set of leaves in $F$, and 
use $|F|$ to denote the number of components in $F$. 
A {\em dangling subtree} of $F$ is the subtree rooted at a vertex of $F$. 

Let $u_1$ and $u_2$ be two vertices in the same component of $F$. 
If $u_1$ and $u_2$ have the same parent in $F$, then they are {\em siblings} in $F$. 
We use $u_1\sim_F u_2$ to denote the path between $u_1$ and $u_2$ in $F$. 
Note that $u_1\sim_F u_2$ is not a directed path if $u_1$ and $u_2$ are incomparable in $F$. 
For convenience, we still 
view each edge of $u_1 \sim_F u_2$ as a directed edge (whose direction is the same as in $F$) 
although $u_1\sim_F u_2$ itself may not be a directed path. Each vertex of $u_1\sim_F u_2$ 
other than $u_1$ and $u_2$ is an {\em inner vertex} of $u_1\sim_F u_2$. 
$D_F(u_1,u_2)$ denotes the set of edges $(p,c)$ in $F$ such that $p$ is an inner vertex of 
$u_1 \sim_F u_2$ but $c$ does not appear in $u_1 \sim_F u_2$. 
Moreover, if $u_1$ and $u_2$ are incomparable in $F$, then 
$D_F^+(u_1,u_2)$ denotes the set consisting of the edges in $D_F(u_1,u_2)$ and 
all defined $e_F(v)$ such that $v$ is a vertex of $u_1 \sim_F u_2$ but 
$e_F(v) \ne e_F(u_i)$ for each $i\in\{1,2\}$ with $u_i \in L(F)$; 
otherwise $D_F^+(u_1,u_2) = \emptyset$. 
For convenience, if $w_1$ and $w_2$ are two vertices in different components in $F$, 
we define $D_F(w_1,w_2) = \emptyset$ and $D_F^+(w_1,w_2) = \emptyset$. 
For example, in Figure~\ref{fig:def}(2), 
$D_F(x_1,x_9) = \{e_F(x_7), e_F(u), e_F(v), e_F(x_3)\}$, while in Figure~\ref{fig:conf}(2), 
$D_{T'}^+(\delta, x_{11})$ consists of the eight dashed edges. 

Let $X$ be a subset of $L(F)$, and $v$ be a vertex of $F$. A descendant $x$ of $v$ in $F$ 
is an {\em $X$-descendant} of $v$ if $x \in X$. 
$X^F(v)$ denotes the set of $X$-descendants of $v$ in $F$. If 
$X^F(v) \ne \emptyset$, then $v$ is {\em $X$-inclusive}; otherwise, $v$ is {\em $X$-exclusive}. 
If $v$ is a non-leaf and both children of $v$ are $X$-inclusive, $v$ is {\em $X$-bifurcate}. 
An edge of $F$ is {\em $X$-inclusive} (respectively, {\em $X$-exclusive}) if its head is 
$X$-inclusive (respectively, $X$-exclusive). For an $X$-bifurcate $v$ in $F$, 
an {\em $X$-child} of $v$ in $F$ is a descendant $u$ of $v$ in $F$ such 
that each edge in $D_F(v,u)$ is $X$-exclusive and either $u \in X$ or $u$ is $X$-bifurcate; 
we also call $v$ the {\em $X$-parent} of $u$ in $F$; note that $v$ has exactly two 
$X$-children in $F$ and we call them {\em $X$-siblings}. In particular, when 
$X=L(F)$, $X$-parent, $X$-children, and $X$-siblings become parent, 
children, and siblings, respectively. For example, if $F$ is as in 
Figure~\ref{fig:conf}(1) and $X=\{x_1,\ldots,x_4\}$, then  
$v$ is $X$-bifurcate and its $X$-children are $x_2$ and $z$. 
$F\myuparrow_X$ denotes the phylogenetic forest obtained from $F$ 
by removing all $X$-exclusive vertices and all vertices without $X$-bifurcate ancestors. 
See Figure~\ref{fig:def} for an example.

\section{Search Trees}\label{sec:search}
A simple way to compute $d(T,F)$ for a TF-pair $(T,F)$ is to build a {\em search tree} $\Gamma$ 
(whose edges each are associated with a set of edges in $F$) 
recursively as follows. If $(T,F) = (\bot, \bot)$, then $\Gamma$ has only one node and we are 
done. So, assume that $(T,F) \ne (\bot, \bot)$. We first construct the root $\rho$ of $\Gamma$. 
To construct the subtrees of $\Gamma$ rooted at the children of $\rho$, we choose a pair $(x_1,x_2)$ 
of sibling leaves in $T$ and distinguish two cases as follows: 

\begin{figure}[htb]
\centerline{\includegraphics{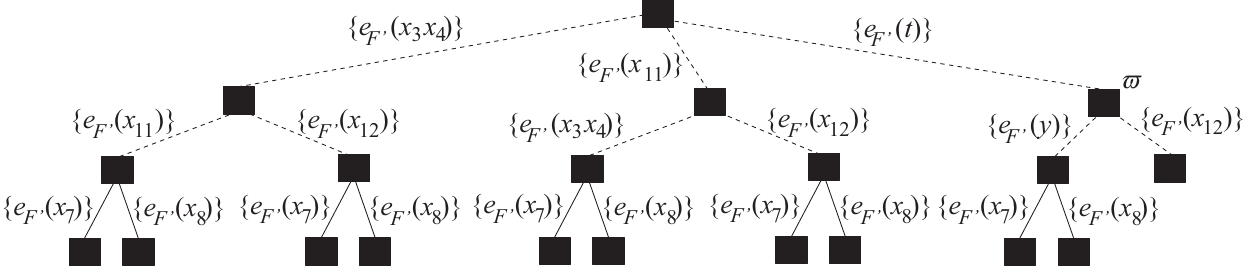}}
\caption{A search tree for the TF-pair $(T',F')$ in Figure~\ref{fig:conf}(2).}
\label{fig:st}
\end{figure}

\medskip
{\em Case 1:} $x_1$ and $x_2$ fall into the same component of $F$. In this case, 
	let $C_1 = \{e_F(x_1)\}$, $C_2 = \{e_F(x_2)\}$, and $C_3 = D_F(x_1,x_2)$. 
	For each $i\in\{1,\ldots,3\}$, we recursively construct a search tree $\Gamma_i$ 
	for the sub-TF pair induced by $C_i$, make the root $\rho_i$ of $\Gamma_i$ 
	be the $i$-th child of $\rho$ by adding the edge $(\rho, \rho_i)$, associate $C_i$  
	with the edge $(\rho, \rho_i)$, and modify the set $C$ associated with each edge 
	in $\Gamma_i$ by replacing each $e \in C$ with the edge of $F$ corresponding to $e$. 

\medskip
{\em Case 2:} $x_1$ and $x_2$ fall into different components of $F$. In this case, 
	$D_F(x_1,x_2)$ is undefined; we construct $\Gamma$ as in Case~1 by ignoring $C_3$, 
	i.e., we do not construct the third child and its descendants. 

\medskip
Note that the children of each node in $\Gamma$ are ordered. 
This finishes the construction of $\Gamma$. See Figure~\ref{fig:st} for an example. 

A path in $\Gamma$ is a {\em root-path} if it starts at the root of $\Gamma$. 
For a root-path $P$ in $\Gamma$, we use $C(P)$ to denote the union of the sets 
associated with the edges in $P$. Clearly, $C(P)$ is a canonical cut of $F$. 
In particular, $C(P)$ is a canonical agreement cut of $(T,F)$ if $P$ ends at 
a leaf node of $\Gamma$. For a node $\varpi$ in $\Gamma$, we use $C(\varpi)$ to denote 
$C(P_{\varpi})$, where $P_{\varpi}$ is the path from the root to $\varpi$ in $\Gamma$. 

Let $\varpi$ be a non-leaf node in $\Gamma$, $C_T$ be the cut of $T$ forced by $C(\varpi)$, 
$(T',F')$ be the sub-TF pair of $(T,F)$ induced by $C(\varpi)$, and $(\alpha_1,\alpha_2)$ be 
the pair of sibling leaves in $T'$ selected to construct the children of $\varpi$ in $\Gamma$. 
The {\em cherry picked at $\varpi$} is $(Y_1,Y_2)$, where for each $i\in\{1,2\}$, 
$Y_i$ is the set of leaf descendants of $\alpha_i$ in $T \ominus C_T$. For example, 
if $(T,F)$ is the same as $(T',F')$ in Figure~\ref{fig:conf}(2), then 
the cherry picked at $\varpi$ is $(\{x_3x_4,x_{11}\}, \{x_{12}\})$. 

A root-path $P$ in $\Gamma$ {\em picks a cherry $(Y_1,Y_2)$} if $P$ contains a node $\varpi$ 
at which $P$ does not end and the cherry picked is $(Y_1,Y_2)$. If in addition, $P$ 
contains the $k$-th child of $\varpi$, then $P$ {\em picks $(Y_1,Y_2)$ in the $k$-th way}. 
Moreover, $P$ {\em isolates} $Y_k$ if $k \in\{1,2\}$, while $P$ {\em creates} $Y_1 \cup Y_2$ 
if $k = 3$. 

$(T,F)$ may have multiple search trees (depending on the order of cherries picked by 
root-paths in a search tree). Nonetheless, it is widely known that for each search tree 
$\Gamma$ of $(T,F)$, $d(T,F) = \min_{\varpi} |C(\varpi)|$, where $\varpi$ ranges over all leaf 
nodes of $\Gamma$ \cite{WBZ10}. Basically, this is true because a search tree 
represents an exhaustive search of a smallest agreement cut of $(T,F)$. 

Let $Z$ be a set of leaves in $T$ such that each component of $T\myuparrow_Z$ is 
a dangling subtree of $T$. A search tree $\Gamma$ of $(T,F)$ {\em respects $Z$}, if 
$P$ can be cut into two portions $P_1$ and $P_2$ such that $P_1$ starts at the root of 
$\Gamma$ and always picks a cherry $(Z_1,Z_2)$ with $Z_1\cup Z_2\subseteq Z$ while $P_2$ 
never picks such a cherry. Since each component of $T\myuparrow_Z$ is a dangling subtree 
of $T$, at least one search tree of $(T,F)$ respects $Z$. For each search tree $\Gamma$ 
of $(T,F)$ respecting $Z$, the tree obtained from $\Gamma$ by deleting all nodes at which 
the cherries $(Z_1,Z_2)$ picked satisfy $Z_1\cup Z_2 \not\subseteq Z$ is called 
a {\em $Z$-search tree} of $(T,F)$. For example, the dashed lines in Figure~\ref{fig:st} form 
a $\{x_3x_4,x_{11},x_{12}\}$-search tree for the TF-pair $(T',F')$ in Figure~\ref{fig:conf}(2).

Let $\Gamma_Z$ be a $Z$-search tree of $(T,F)$. A {\em root-leaf path} in $\Gamma_Z$ 
is a root-path $P$ in $\Gamma_Z$ ending at a leaf node of $\Gamma_Z$. Since each $Z$-search 
tree $\Gamma_Z$ of $(T,F)$ can be extended to a search tree of $(T,F)$, the following hold: 

\begin{itemize}
\item For each root-leaf path $P$ in $\Gamma_Z$, $C(P)$ is a canonical cut of $F$ 
	and can be extended to a canonical agreement cut of $(T,F)$. 
\item $\Gamma_Z$ has at least one root-leaf path $P$ such that $C(P)$ can be extended 
	to a smallest canonical agreement cut of $(T,F)$. 
\end{itemize}

\section{Keys and Lower Bounds}\label{sec:key}
Throughout this section, let $(T,F)$ be a TF pair. Instead of cuts, we consider a more useful 
notion of {\em key}. Intuitively speaking, a key contains not only a cut $B$ within a local 
structure of $F$ but also possibly two leaves in $F\ominus B$ to be merged into a single leaf. 
Formally, a {\em key} of $(T,F)$ is a triple $\kappa=(X, B, R)$ satisfying the following conditions:
\begin{enumerate}
\item\label{cond:dangle} 
	$X$ is a set of leaves in $T$ such that each component of $T\myuparrow_X$ is 
	a dangling subtree of $T$.  
\item $B \subseteq E_X$ is a cut of $F$, where $E_X$ is the set of edges in $F$ 
whose tails or heads appear in $F\myuparrow_X$. 
Moreover, either $\{e_F(x)~|~x\in X\} \subseteq B$ or $|\{e_F(x)~|~x\in X\} \setminus B| = 2$. 
\item\label{cond:R} 
If $\{e_F(x)~|~x\in X\} \subseteq B$, then $R=\emptyset$; otherwise, for the two vertices $x_1$ 
and $x_2$ in $X$ with $\{e_F(x_1), e_F(x_2)\} = \{e_F(x)~|~x\in X\} \setminus B$, we have that 
$x_1$ and $x_2$ are siblings in $F\ominus B$, $R$ is the edge set of $x_1 \sim_F x_2$, 
and $B \cap R = \emptyset$. ({\em Comment:} By Condition~\ref{cond:dangle}, $x_1$ and $x_2$ are 
siblings in $T\ominus \{e_T(x)~|~x\in X\setminus\{x_1,x_2\}\}$ as well. So, when we compute 
the sub-TF pair of $(T,F)$ induced by $B$, $x_1\sim_F x_2$ will be contracted into 
a single leaf and so will $x_1\sim_T x_2$.) 
\end{enumerate}
For example, if $(T,F)$ is as in Figure~\ref{fig:conf}(1), then $\kappa_e=(X,B,R)$ is 
a key of $(T,F)$, where $X=\{x_2,x_3,x_4\}$, $B=\{e_F(x_1), e_F(x_2), e_F(u)\}$, 
and $R$ is the edge set of $x_3\sim_F x_4$. 

If $R = \emptyset$, then $\kappa$ is {\em normal} and we simply write $\kappa=(X,B)$ 
instead of $\kappa=(X,B,R)$; otherwise, it is {\em abnormal}. 
In essence, only normal keys were considered in \cite{Shi+14}. 

Let $\kappa=(X,B,R)$ be a key of $(T,F)$. The {\em size} of $\kappa$ is $|B|$ and is 
also denoted by $|\kappa|$. The {\em sub-TF pair of $(T,F)$ induced by $\kappa$} is 
the sub-TF pair $(T',F')$ of $(T, F)$ induced by $B$. Let $Y$ be a subset of $L(T)$ 
such that $X\subseteq Y$ and each component in $T\myuparrow_Y$ is a dangling subtree 
of $T$. Further let $P$ be a root-leaf path in a $Y$-search tree of $(T,F)$. A component 
$K$ in $F - ((C(P)\setminus R) \cup B)$ is {\em free} if $K$ contains no labeled leaf. 
We use $f_c(\kappa,P)$ to denote the set of free components in $F - ((C(P)\setminus R) 
\cup B)$, and use $f_e(\kappa,P)$ to denote $C(P)\cap (B \cup R)$.  
The {\em lower bound} achieved by $\kappa$ $\wrt$ $P$ is $b(\kappa,P) = 
|f_e(\kappa,P)| + |f_c(\kappa,P)|$. 

The {\em lower bound} achieved by $\kappa$ via $Y$ is $b_Y(\kappa)= \displaystyle 
\max_{\Gamma}\min_P b(\kappa,P)$, where $\Gamma$ ranges over all $Y$-search trees 
of $(T,F)$ and $P$ ranges over all root-leaf paths in $\Gamma$. If $\Gamma$ is a 
$Y$-search tree of $(T,F)$ with $b_Y(\kappa) = \min_P b(\kappa,P)$ where $P$ ranges 
all root-leaf paths in $\Gamma$, then $\Gamma$ is called a $Y$-search tree of $(T,F)$ 
{\em witnessing} $b_Y(\kappa)$. 
When $Y = X$, we write $b(\kappa)$ instead of $b_X(\kappa)$. 
The next two lemmas may help the reader understand the definitions 
(especially for abnormal keys) and will also be useful later. 

\begin{lemma}\label{lem:size2} Suppose that $T$ has a pair $(x_1,x_2)$ of sibling 
leaves such that $\ell_F(x_1,x_2)$ exists and $1\le |D_F(x_1,x_2)| \le 2$. Then, 
$(T,F)$ has an abnormal key $\kappa$ with $|\kappa| \le 2$ and $b(\kappa) \ge 1$.
\end{lemma}
\begin{proof}
Let $\kappa = (X,B,R)$, where $X = \{x_1,x_2\}$, $B = D_F(x_1,x_2)$, and $R$ is the edge set 
of $x_1 \sim_F x_2$. A root-leaf path $P$ in the unique $X$-search tree of $(T,F)$ either 
creates $X$ or isolates one of $\{x_1\}$ and $\{x_2\}$. In the former case, $B\subseteq C(P)$ 
and in turn $|C(P)\cap B|\ge 1$. In the latter case, $e_F(x_1) \in C(P)$ 
or $e_F(x_2) \in C(P)$, and hence $|C(P)\cap R| \ge 1$. 
\end{proof}
\begin{lemma}\label{lem:2cross}
Suppose that $T$ has two pairs $(x_1,x_2)$ and $(x_3,x_4)$ of sibling leaves such that 
$x_1$ and $x_3$ are siblings in $F$ and so are $x_2$ and $x_4$. Then, $(T,F)$ has 
a normal key $\hat{\kappa} = (\{x_1,\ldots,x_4\}, \hat{B})$ with 
$|\hat{\kappa}| \le 2 b(\hat{\kappa})$.
\end{lemma}
\begin{proof}
Let $\hat{X}=\{x_1,\ldots,x_4\}$ and $\hat{B}=\{e_F(x_1),\ldots,e_F(x_4)\}$. 
Consider an $X$-search tree $\Gamma$ of $(T,F)$ in which each root-leaf path $P$ first 
picks the cherry $(\{x_1\}, \{x_2\})$ and then picks the cherry $(\{x_3\}, \{x_4\})$. 
If for some $i\in\{1,3\}$, $P$ creates $\{x_i, x_{i+1}\}$, then 
$C(P)$ contains both $e_F(x_{4-i})$ and $e_F(x_{5-i})$. Otherwise, 
$C(P)$ contains $e_F(x_i)$ and $e_F(x_j)$ for some $i\in\{1,2\}$ and $j\in\{3,4\}$. 
So, in any case, $|C(P)\cap B| = 2$. 
\end{proof}

For the next three lemmas, let $\kappa=(X,B,R)$ be a key of $(T,F)$. 
The lemmas show why we can call $b_{L(T)}(\kappa)$ a lower bound. 
The next lemma is obvious. 

\begin{lemma}\label{lem:f_c1}
Let $P$ be a root-leaf path in an $X$-search tree of $(T,F)$, and 
$H$ be a component in $f_c(\kappa,P)$. Then, there is no $e\in R$ such that 
some endpoint of $e$ is in $H$. Moreover, either the root $r_H$ of $H$ is a root in $F$, 
or $(C(P)\setminus R) \cup B$ contains $e_F(r_H)$.  
\end{lemma}

\begin{lemma}\label{lem:ub0}
Let $Y$ be a subset of $L(T)$ such that $X\subseteq Y$ and each component in 
$T\myuparrow_Y$ is a dangling subtree of $T$. Then, $b_Y(\kappa) \le b_{L(T)}(\kappa)$. 
\end{lemma}
\begin{proof}
Let $\Gamma_Y$ be a $Y$-search tree of $(T,F)$ witnessing $b_Y(\kappa)$. 
As aforementioned, we can extend $\Gamma_Y$ to a search tree $\Gamma$ of $(T,F)$. 
In other words, each root-leaf path $P_Y$ in $\Gamma_Y$ is the first portion of 
a root-leaf path $P$ in $\Gamma$. Since $C(P_Y) \subseteq C(P)$, $C(P_Y) \cap (B\cup R) 
\subseteq C(P) \cap (B\cup R)$, and $|f_c(\kappa,P_Y)| \le |f_c(\kappa,P)|$. 
In summary, $b(\kappa,P_Y) \le b(\kappa,P)$. 
\end{proof}

The next lemma shows why $b(\kappa)$ is a lower bound on $d(T,F) - d(T',F')$. 

\begin{lemma}\label{lem:ub}
Let $P$ be a root-leaf path in a search tree of $(T,F)$ such that 
$C(P)$ is a smallest canonical agreement cut of $(T,F)$. Then, 
$d(T,F) - d(T',F') \ge b(\kappa,P)$, where $(T',F')$ is the sub-TF pair of $(T,F)$ 
induced by $\kappa$. Consequently, $d(T,F) - d(T',F') \ge b_{L(T)}(\kappa) \ge b(\kappa)$. 
\end{lemma}
\begin{proof}
Let $M = C(P)\setminus R$. Obviously, $|M| = |C(P)| - |C(P)\cap R| = d(T,F) - |C(P)\cap R|$ 
and $F - M$ has $|F| + d(T,F) - |C(P)\cap R|$ components. Since $C(P)$ is a canonical cut 
of $F$, every component in $F - C(P)$ has a labeled leaf. So, $F \ominus M$ has the same number 
of components as $F - M$. Moreover, $|B \setminus M| = |B| - |C(P)\cap B|$ and in turn 
$F - (M \cup B)$ has exactly $|F\ominus M| + |B| - |C(P)\cap B|$ components among which 
$f_c(\kappa,P)$ have no labeled leaves. Hence, $|F \ominus (M \cup B)| \le \left(d(T,F) + 
|F| - |C(P)\cap R|\right) + \left(|B| - |C(P)\cap B| - |f_c(\kappa,P)|\right)
= |F| + |B| + d(T,F) - b(\kappa,P)$. 

Since $F \ominus C(P)$ is an agreement forest of $(T,F)$, $F \ominus (C(P) \cup B)$ is clearly 
an agreement forest of $(T,F \ominus B)$. So, by Condition~\ref{cond:R} in the definition 
of keys, $F \ominus (M \cup B)$ is also an agreement forest of $(T, F\ominus B)$. 
Now, since $|F \ominus B| = |F| + |B|$, $(T, F\ominus B)$ has an agreement cut of size 
at most $d(T,F) - b(\kappa,P)$. On the other hand, a smallest agreement cut of 
$(T,F \ominus B)$ is of size $d(T',F')$. Hence, $d(T,F) - d(T',F') \ge b(\kappa,P)$. 
Therefore, $d(T,F) - d(T',F') \ge b_{L(T)}(\kappa)$, and in turn 
$d(T,F) - d(T',F') \ge b(\kappa)$ by Lemma~\ref{lem:ub0}. 
\end{proof}

A key $\kappa$ of $(T,F)$ is {\em good} if $|\kappa| \le 2b_{L(T)}(\kappa)$, while 
$\kappa$ is {\em fair} if $|\kappa| \le 2b_{L(T)}(\kappa) + 1$). 
Obviously, for each leaf $x$ of $T$, $(\{x\}, \{e_F(x)\})$ is a fair key of $(T,F)$. 
If $\kappa=(X,B,R)$ is a good key of $(T,F)$, then by Lemma~\ref{lem:ub}, 
$B$ is a good cut of $F$. So, in order to find a good cut of $F$, 
it suffices to find a good key of $(T,F)$. 

The next three lemmas can often help us estimate $f_e(\kappa,P)$ and $f_c(\kappa,P)$. 
For the three lemmas, let $\kappa=(X,B)$ be a normal key of $(T,F)$, and $P$ be 
a root-leaf path in an $X$-search tree $\Gamma_X$ of $(T,F)$ witnessing $b(\kappa)$. 

\begin{lemma}\label{lem:cutHelp}
Suppose that $(\varpi_1,\varpi_2)$ is an edge in $P$ such that some edge $e_F(v) \in 
C(\varpi_2) \setminus C(\varpi_1)$ satisfies that the set $Y$ of labeled-leaf descendants 
of $v$ in $F - C(\varpi_1)$ is a subset of $X$. Then, the following hold:
\begin{enumerate}
\item If $|Y| \ge 2$, then $v$ is the root of a component in $f_c(\kappa,P)$. 
\item If $|Y| = 1$, then either $Y=\{v\}$ and $e_F(v) \in f_e(\kappa,P)$, 
or $Y\ne\{v\}$ and $v$ is the root of a component in $f_c(\kappa,P)$. 
\end{enumerate}
\end{lemma}
\begin{proof}
Since $C(\varpi_2)$ is a canonical cut of $F$, $F - C(\varpi_2)$ has a component $K$ 
whose leaf set is $Y$. The root of $K$ is $v$. Obviously, if $v \in X$, then $Y=\{v\}$ 
and $e_F(v) \in C(P) \cap B$. Otherwise, the leaves of $K$ are the tails of the edges 
in $\{e_F(x)~|~x\in Y\}$ and hence are all unlabeled; consequently, the leaf descendants 
of $v$ in $F - (C(P) \cup B)$ are all unlabeled and in turn $f_c(\kappa,P)$ contains a 
component rooted at $v$.  
\end{proof}

\begin{lemma}\label{lem:mergeHelp}
Suppose that $(\varpi_1,\varpi_2)$ is an edge in $P$ such that $P$ picks some cherry 
$(Y_1,Y_2)$ at $\varpi_1$ by creating $Y_1 \cup Y_2$. Let $u_1=\ell_F(Y_1)$, 
$u_2=\ell_F(Y_2)$, and $Y = Y_1\cup Y_2$. Then, the following hold:
\begin{enumerate}
\item If $\ell_F(Y)$ is a root of $F$, then it is also the root of a component in $f_c(\kappa,P)$. 
\item $B\cap D_F(u_1, u_2) \subseteq f_e(\kappa,P)$. 
\item For each $e=(v,w)\in B\cap \left(D^+_F(u_1, u_2) \setminus D_F(u_1, u_2)\right)$, 
$w$ is the root of a component in $f_c(\kappa,P)$. 
\end{enumerate} 
\end{lemma}
\begin{proof}
Let $C_T$ be the cut of $T$ forced by $C(\varpi_2)$, $T' = T \ominus C_T$, and 
$F' = F \ominus C(\varpi_2)$. Since $P$ creates $Y_1\cup Y_2$, $\ell_{T'}(Y)$ 
and $\ell_{F'}(Y)$ agree and in turn $F\myuparrow_Y$ is a dangling subtree of $F - C(P)$. 
Thus, Statement~1 holds. Statement~2 holds because $D_F(u_1, u_2) \subseteq C(P)$. 
To prove Statement~3, consider an edge $e=(v,w)\in B\cap \left(D_F^+(u_1,u_2) \setminus 
D_F(u_1,u_2)\right)$. Note that either $e$ appears in $u_1 \sim_F u_2$ or $w$ is the root 
of $F\myuparrow_Y$. In either case, since $C(P)$ is a canonical cut of $F$, each leaf 
descendant of $w$ in $F - C(P)$ belongs to $Y$. Now, since $\{e_F(x)~|~x\in Y\}\subseteq B$, 
the component of $F - (C(P)\cup B)$ rooted 
at $w$ has no labeled leaf and hence belongs to $f_c(\kappa,P)$. 
\end{proof}

\begin{lemma}\label{lem:f_c2}
Suppose $\tau=(X,\hat{B})$ is another normal key of $(T,F)$ such that 
$B\subseteq \hat{B}$ and $|\hat{B}\setminus B| = 1$. Then, $f_e(\kappa,P) 
\subseteq f_e(\tau,P)$. Moreover, either $f_c(\kappa,P) \subseteq f_c(\tau,P)$, or 
$f_c(\kappa,P) \not\subseteq f_c(\tau,P)$ and $|f_c(\tau,P)| = |f_c(\kappa,P)| + 1$.
\end{lemma}
\begin{proof}
Since $B\subseteq \hat{B}$, $f_e(\kappa,P) \subseteq f_e(\tau,P)$. 
Let $\hat{e}$ be the unique edge in $\hat{B}\setminus B$. 
If no component in $f_c(\kappa,P)$ contains $\hat{e}$, then clearly 
$f_c(\kappa,P) \subseteq f_c(\tau,P)$. Otherwise, $f_c(\tau,P)$ can be obtained 
from $f_c(\kappa,P)$ by splitting the component containing $\hat{e}$ into two 
components neither of which contains a labeled leaf. 
\end{proof}

\section{Finding a Good Cut}\label{sec:goodKey}
In this section, we prove Theorem~\ref{th:goodRatio}. Hereafter, 
fix a TF-pair $(T,F)$ such that for each pair $(x_1,x_2)$ of 
sibling leaves in $T$, $D_F(x_1,x_2) \ne \emptyset$. For a vertex $\alpha$ in $T$, 
we use $L_{\alpha}$ to denote the set of leaf descendants of $\alpha$ in $T$. 

To find a good key of $(T,F)$, our basic strategy is to process $T$ in a bottom-up fashion. 
At the bottom (i.e., when processing a leaf $x$ of $T$), we can easily construct a fair 
key $(\{x\}, \{e_F(x)\})$ of $(T,F)$. For a non-leaf $\alpha$ of $T$ 
with children $\alpha_1$ and $\alpha_2$ in $T$ such that a fair key $\kappa_i = 
(L_{\alpha_i}, B_i)$ has been constructed for each $i\in\{1,2\}$, we want to 
combine $\kappa_1$ and $\kappa_2$ into a fair key $\kappa = (L_\alpha, B)$ with 
$B_1 \cup B_2 \subseteq B$. As one can expect, this can be done only when $\kappa_1$, 
$\kappa_2$, and $\alpha$ satisfy certain conditions. 
So, we first figure out the conditions below.

A vertex $\alpha$ of $T$ is {\em consistent} with $F$ if either $\alpha$ is a leaf, or $\alpha$ 
is a non-leaf in $T$ such that $F\myuparrow_{L_\alpha}$ is a tree and the root in 
$T\myuparrow_{L_\alpha}$ agrees with the root in the binarization of 
$F\myuparrow_{L_\alpha}$. For example, if $(T,F)$ is as in Figure~\ref{fig:conf}, 
then $\beta$ is consistent with $F$ but $\alpha$ is not. 

\begin{lemma}\label{lem:consistent}
Suppose that a vertex $\alpha$ of $T$ is consistent with $F$. Let $P$ be a root-leaf path 
in an $L_\alpha$-search tree of $(T,F)$, and $C_T$ be the cut of $T$ forced by $C(P)$. 
Then, the following hold:
\begin{enumerate}
\item Let $T'=T \ominus C_T$, $F'=F \ominus C(P)$, and $Y$ be the set of leaf descendants 
of $\alpha$ in $T - C_T$. Then, $\ell_{T'}(Y)$ and $\ell_{F'}(Y)$ agree. 

\item The tail of each edge in $C_T$ (respectively, $C(P)$) is a proper descendant 
of $\alpha$ (respectively, $\ell_F(L_\alpha)$) in $T$ (respectively, $F$). 

\item If $\ell_F(\ell_\alpha)$ is not a root in $F$, then for every normal key 
$\kappa=(L_\alpha,B)$ of $(T,F)$, the root $r_H$ of each component $H$ in 
$f_c(\kappa,P)$ is a descendant of $\ell_F(L_\alpha)$ in $F$. 
\end{enumerate}
\end{lemma}
\begin{proof}
Statements~1 and~2 can be trivially proved via induction on $|L_\alpha|$. 
Statement~3 follows from Statement~2 and Lemma~\ref{lem:f_c1} immediately. 
\end{proof}

A {\em robust key} of $(T,F)$ is a normal key $\kappa=(X,B)$ such that $\ell_T(X)$ is 
consistent with $F$, $\ell_F(X)$ is not a root of $F$, $e_F(X) \not\in B$, 
and $\ell_F(X)$ has a labeled descendant in $F - B$. If in addition, $\ell_F(X)$ has 
two children in $F-B$ and each of them has a labeled descendant in $F - B$, then $\kappa$ 
is {\em super-robust}. For example, if $T$ and $F$ are as in Figure~\ref{fig:conf}(1), then 
$\kappa=(\{x_2,\ldots,x_4\},\{e_F(x_1), \ldots, e_F(x_4), e_F(x_7)\})$ is a robust but not 
super-robust key of $(T,F)$. 

The next lemma shows how to combine two normal keys into one larger normal key. 

\begin{lemma}\label{lem:helpNormal}
Let $\alpha$ be a non-leaf in $T$, and $\beta_1$ and $\beta_2$ be its children in $T$. 
Suppose that both ${\beta_1}$ and $\beta_2$ are consistent with $F$ and 
neither $\ell_F(L_{\beta_1})$ nor $\ell_F(L_{\beta_2})$ is a root of $F$. 
Further assume that $(T,F)$ has two fair normal keys $\kappa_1 = (L_{\beta_1},B_1)$ 
and $\kappa_2 = (L_{\beta_2},B_2)$. Then, the following hold: 

\begin{enumerate}
\item If either $\ell_F(L_\alpha)$ is undefined, or $\alpha$ is consistent with $F$ 
and $\ell_F(L_\alpha)$ is a root of $F$, then $\kappa = (L_\alpha, B_1\cup B_2)$ is 
a good normal key of $(T,F)$.

\item If $\alpha$ is consistent with $F$, $\ell_F(L_\alpha)$ is not a root of $F$, 
and $|D_F\left( \ell_F(L_{\beta_1}) , \ell_F(L_{\beta_2}) \right)| + t \ge 1$ where 
$t$ is the number of robust keys among $\kappa_1$ and $\kappa_2$, then 
	$\kappa = (L_\alpha, B_1\cup B_2 \cup \{\tilde{e}\})$ is a fair normal key of $(T,F)$, 
	where $\tilde{e}$ is an arbitrary edge in $D_F^+(\ell_F(L_{\beta_1}),\ell_F(L_{\beta_2})) 
	\setminus \{e_F(L_{\beta_i})~|~1\le i\le 2$ and $\kappa_i$ is not robust$\}$ or any edge 
	entering a child of $\ell_F(L_{\beta_i})$ in $F$ for some super-robust $\kappa_i$ 
	with $i\in\{1,2\}$. 
\end{enumerate}
\end{lemma}
\begin{proof}
Obviously, $\kappa$ is a normal key of $(T,F)$. 
It remains to show that $b(\kappa) \ge b(\kappa_1) + b(\kappa_2) + 1$.

For each $i\in\{1,2\}$, let $\Gamma_{{i}}$ be an $L_{\beta_i}$-search tree witnessing $b(\kappa_i)$. 
Since $\alpha$ is consistent with $F$, 
we can combine $\Gamma_{{1}}$ and $\Gamma_{{2}}$ into an $L_\alpha$-search tree $\Gamma$ 
of $(T,F)$ such that each root-leaf path $P$ in $\Gamma$ can be cut into three portions 
$P_1$, $P_2$, $Q$, where $P_i$ corresponds to a root-leaf path in $\Gamma_{{i}}$ for each 
$i\in\{1,2\}$, while $Q$ is a single edge corresponding to picking a cherry $(Y_1,Y_2)$ with 
$Y_1\subseteq L_{\beta_1}$ and $Y_2\subseteq L_{\beta_2}$. By Statement~2 in Lemma~\ref{lem:consistent}, 
$B_1 \cap B_2 = \emptyset$ and in turn $f_e(\kappa_1, P_1) \cap f_e(\kappa_2, P_2) = \emptyset$. 
Moreover, since $B$ is a superset of both $B_1$ and $B_2$, 
$f_e(\kappa_1, P_1) \cup f_e(\kappa_2, P_2) \subseteq f_e(\kappa, P)$. Furthermore, 
Statement~3 in Lemma~\ref{lem:consistent} ensures that 
$f_c(\kappa_1, P_1) \cap f_c(\kappa_2, P_2) = \emptyset$ and 
$f_c(\kappa_1, P_1) \cup f_c(\kappa_2, P_2) \subseteq f_c(\kappa, P)$. 

For convenience, let $F' = F - (C(P_1)\cup C(P_2))$. By Statements~1 and 2 in Lemma~\ref{lem:consistent}, 
$F\myuparrow_{Y_i}$ is a dangling subtree of $F'$ and the head of $e_{F'}(Y_i)$ 
is a descendant of the head of $e_F(L_{\beta_i})$. If $Q$ isolates $Y_1$ or $Y_2$, then 
by Lemma~\ref{lem:cutHelp}, $b(\kappa) \ge b(\kappa,P) \ge b(\kappa_1,P_1) + b(\kappa_2,P_2) + 1 
\ge b(\kappa_1) + b(\kappa_2) + 1 $. So, suppose that $Q$ creates $Y_1 \cup Y_2$. Then, by 
Lemma~\ref{lem:mergeHelp}, we cannot guarantee $b(\kappa,P) \ge b(\kappa_1,P_1) + b(\kappa_2,P_2) + 1$ 
only if $D^+_F(\ell_F(Y_1), \ell_F(Y_2)) \cap (B\setminus(B_1\cup B_2)) = \emptyset$. Since $\tilde{e} 
\in B\setminus(B_1\cup B_2)$, $D^+_F(\ell_F(Y_1), \ell_F(Y_2)) \cap (B\setminus(B_1\cup B_2)) = 
\emptyset$ can happen only if for some $i\in\{1,2\}$, $\tilde{e}$ is the edge entering a child of 
$\ell_F(L_{\beta_i})$ in $F$ and $\ell_F(L_{\beta_i}) = \ell_F(Y_i)$. Thus, we may further assume that 
such an $i$ exists. Then, $P_i$ picks its last cherry in the third way and in turn $\tilde{e} \in 
f_e(\kappa,P)$ by Lemma~\ref{lem:mergeHelp}. 
\end{proof}

We may have two or more choices for $\tilde{e}$ in Lemma~\ref{lem:helpNormal}. 
In that case, sometimes we may want to choose $\tilde{e}$ (in the listed order) as follows: 
\begin{itemize}
\item If for some $i\in\{1,2\}$, either $|D_F\left( \ell_F(L_{\beta_i}), \ell_F(L_{\alpha}) \right)| \ge 2$, 
	or $|D_F\left( \ell_F(L_{\beta_i}), \ell_F(L_\alpha) \right)| = 1$ and $\kappa_i$ is robust, then 
	$\tilde{e}$ is an arbitrary edge in $D_F\left( \ell_F(L_{\beta_i}), \ell_F(L_\alpha) \right)$. 
\item If $\kappa_i$ is super-robust for some $i\in\{1,2\}$, then 
	$\tilde{e}$ is any edge entering a child of $\ell_F(L_{\beta_i})$ in $F$. 
\item If $D_F\left( \ell_F(L_{\beta_1}), \ell_F(L_{\beta_2}) \right) \ne \emptyset$, then 
	$\tilde{e}$ is an arbitrary edge in it; otherwise, $\tilde{e} = e_F(L_{\beta_i})$ 
	for an arbitrary $i\in\{1,2\}$ such that $\kappa_i$ is robust. 
\end{itemize}
We refer to the above way of choosing $\tilde{e}$ as the {\em robust way} of combining 
$\kappa_1$ and $\kappa_2$. Intuitively speaking, we try to make $\kappa$ super-robust; 
if we fail, we then try to make $\kappa$ robust. For example, 
if $|D_F\left( \ell_F(L_{\beta_1}) , \ell_FL_{\beta_2}) \right)| + t \ge 2$ in 
Lemma~\ref{lem:helpNormal}, choosing $\tilde{e}$ via the robust way 
yields a robust $\kappa$. 

Obviously, for each leaf $x$ of $T$, $\kappa = (\{x\}, \{e_F(x)\})$ is a normal key of 
$(T,F)$ with $|\kappa| = 1$ and $b(\kappa) \ge 0$. Thus, the next lemma follows from 
Statements~2 in Lemma~\ref{lem:helpNormal} immediately. 

\begin{lemma}\label{lem:trivial2} Suppose that $T$ has a pair $(x_1,x_2)$ of sibling leaves 
such that $|D_F(x_1,x_2)| \ge 2$ and $\ell_F(x_1,x_2)$ is not a root of $F$. Then, for every 
$e \in D_F(x_1,x_2)$, $\kappa=(\{x_1,x_2\},\{e_F(x_1), e_F(x_2), e\})$ is a fair robust 
key of $(T,F)$; if in addition $|D_F(x_1,x_2)| \ge 3$, neither child of $\ell_F(x_1,x_2)$ 
in $F$ is a leaf, and $e \in D_F(x_i,\ell_F(x_1,x_2))$ for some $i\in\{1,2\}$ with 
$|D_F(x_i,\ell_F(x_1,x_2))| \ge 2$, then $\kappa$ is super-robust. 
\end{lemma}

Basically, Lemma~\ref{lem:helpNormal} tells us the following. After processing $\beta_1$ and 
$\beta_2$, we can move up in $T$ to process $\alpha$ (as long as $\alpha$ satisfies certain
conditions). However, the condition $|D_F\left( \ell_F(L_{\beta_1}) , \ell_F(L_{\beta_2}) 
\right)| + t \ge 1$ in Lemma~\ref{lem:helpNormal} is not so easy to use. 
We next figure out an easier-to-use condition. 

Let $A$ be a (possibly empty) set of edges in $F$, and $X$ be a subset of $L(T)$. 
An {\em $X$-path} in $F-A$ is a directed path $q$ to an $x\in X$ in $F-A$ such that 
each vertex of $q$ that is bifurcate in $F-A$ is also $X$-bifurcate in $F-A$. 
For each vertex $v$ of $F-A$, let $N_{A,X}(v)$ denote the number of $X$-paths starting 
at $v$ in $F-A$. When $A=\emptyset$, we write $N_{X}(v)$ instead of $N_{A,X}(v)$. 

\medskip
{\bf Example:} Let $T$ and $F$ be as in Figure~\ref{fig:conf}(1), and 
$X=\{x_2,x_3,x_4\}$. Then, $w\sim_F x_4$ is an $X$-path in $F$ but 
$v\sim_F x_3$ is not; indeed, $N_X(v) = 0$, $N_X(w) = 1$. 
However, if $A=\{e_F(x_1)\}$, then $v\sim_F x_3$ is an $X$-path in $F-A$, 
$N_{A,X}(v) = 1$, and $N_{A,X}(w) = 2$. 

\begin{lemma}\label{lem:goUp}
Let $\alpha$ be a vertex of $T$ consistent with $F$ such that $\ell_F(L_\alpha)$ is not a root of $F$ 
and $N_{L_\alpha}(v) \le 1$ for each descendant $v$ of $\ell_F(L_\alpha)$ in $F$. Then, $(T,F)$ has 
a fair normal key $\kappa = (L_\alpha, B)$ such that $\kappa$ is robust 
if $N_{L_\alpha}({\ell_F({L_\alpha})}) = 0$. 
\end{lemma}
\begin{proof}
By induction on $|{L_\alpha}|$. The lemma is clearly true when $\alpha$ is a leaf of $T$. 
So, suppose that $\alpha$ is a non-leaf of $T$. Let $\beta_1$ (respectively, $\beta_2$) be 
the children of $\alpha$ in $T$. For each $i\in\{1,2\}$, let $\kappa_i = (L_{\beta_i}, B_i)$ be 
a normal key of $(T,F)$ guaranteed by the inductive hypothesis for $\beta_i$. Let $t$ be defined as 
in Lemma~\ref{lem:helpNormal}. 

\medskip
{\em Case 1:} $N_{L_\alpha}({\ell_F({L_\alpha})}) = 1$. In this case, there is exactly one $i\in\{1,2\}$ 
with $N_{L_\alpha}(\ell_F(L_{\beta_i})) = 1$. We may assume $i=1$. Then, $D_F(\ell_F(L_{\alpha}),\ell_F(L_{\beta_2})) 
\ne \emptyset$, or $N_{L_\alpha}(\ell_F(L_{\beta_2})) = 0$ 
and in turn $\kappa_2$ is robust by the inductive hypothesis. 
In either case, Statement~2 in Lemma~\ref{lem:helpNormal} ensures that 
$\kappa = (L_\alpha, B)$ with $B=B_1\cup B_2 \cup \{e_F(L_\alpha)\}$ satisfies the conditions in the lemma. 

\medskip
{\em Case 2:} $N_{L_\alpha}({\ell_F(L_\alpha)}) = 0$. In this case, we claim that 
$|D_F(\ell_F(L_{\beta_1}),\ell_F(L_{\beta_2}))| + t \ge 2$. 
The claim is clearly true if $|D_F(\ell_F(L_{\beta_1}),\ell_F(L_{\beta_2}))| \ge 2$. 
Moreover, if $D_F(\ell_F(L_{\beta_1}),\ell_F(L_{\beta_2})) = \emptyset$, 
then $N_{L_\alpha}(\ell_F(L_\alpha)) = N_{L_\alpha}(\ell_F(L_{\beta_1})) + N_{L_\alpha}(\ell_F(L_{\beta_2}))$ and 
in turn $N_{L_\alpha}(\ell_F(L_{\beta_1})) = N_{L_\alpha}(\ell_F(L_{\beta_2})) = 0$; hence, $t \ge 2$ 
by the inductive hypothesis, and the claim holds. So, we assume $|D_F(\ell_F(L_{\beta_1}),\ell_F(L_{\beta_2}))|$ $ = 1$. 
Then, for some $i\in\{1,2\}$, $|D_F(\ell_F(L_{\alpha}),\ell_F(L_{\beta_i}))| = 1$ and 
$D_F(\ell_F(L_{\alpha}),\ell_F(L_{\beta_{3-i}})) = \emptyset$. We may assume  
$i=1$. Then, $N_{L_\alpha}(\ell_F(L_{\beta_2})) = 0$ and in turn $\kappa_2$ is robust by the inductive hypothesis. 
Thus, the claim holds. By the claim, if we set $\tilde{e}$ to be an arbitrary edge in 
$D_F(\ell_F(L_{\beta_1}),\ell_F(L_{\beta_2}))$, or set $\tilde{e} = e_F(L_{\beta_i})$ 
for an arbitrary robust $\kappa_i$, 
then Statement~2 in Lemma~\ref{lem:helpNormal} ensures that $\kappa = (L_\alpha, B)$ with 
$B=B_1\cup B_2 \cup \{\tilde{e}\}$ satisfies the conditions in the lemma. 
\end{proof}

\subsection{The Easy Cases}\label{subsec:easy}
With Lemmas~~\ref{lem:helpNormal} and~\ref{lem:goUp}, we are now ready to state the easy 
cases where we can end up with a good key. Throughout this subsection, let $\alpha$ be 
a vertex of $T$. 

$\alpha$ is a {\em close stopper} for $(T,F)$ 
if $\alpha$ is consistent with $F$, $N_{L_\alpha}({\ell_F({L_\alpha})}) \ge 2$, and 
$N_{L_\alpha}(v) \le 1$ for all proper descendants $v$ of $\ell_F({L_\alpha})$ in $F$. $\alpha$ 
is a {\em semi-close stopper} for $(T,F)$ if $\alpha$ is consistent with $F$ and $L_\alpha$ 
contains $x_1$ and $x_2$ with $\ell_F(x_1,x_2) = \ell_F(L_\alpha)$ such that all but at most 
two edges in $D_F(x_1,x_2)$ are $L_\alpha$-inclusive and the head of each $L_\alpha$-inclusive 
edge in $D_F(x_1,x_2)$ has no descendant $v$ in $F$ with $N_{L_\alpha}(v) \ge 2$. Note that 
a close stopper for $(T,F)$ is also a semi-close stopper for $(T,F)$. 
For example, if $T$ and $F$ are as in Figure~\ref{fig:conf}(1), then $\beta$ is 
a semi-close (but not close) stopper for $(T,F)$, while $\gamma$ is not. 

$\alpha$ is a {\em root stopper} for $(T,F)$ if it is consistent with $F$, no descendant of 
$\alpha$ in $T$ is a semi-close stopper for $(T,F)$, and $\ell_F(L_\alpha)$ is a root in $F$, 
For example, if $T$ and $F$ are as in Figure~\ref{fig:conf}(1), then $\lambda$ looks like a root 
stopper for $(T,F)$ at first glance, but it is not because it is a semi-close stopper for $(T,F)$. 

$\alpha$ is a {\em disconnected stopper} for $(T,F)$ if $\ell_F(L_\alpha)$ is undefined, 
no descendant of $\alpha$ in $T$ is a semi-close or root stopper for $(T,F)$, 
and both children of $\alpha$ in $T$ are consistent with $F$. For example, if $T$ and $F$ are 
as in Figure~\ref{fig:conf}(1), then $\zeta$ is a disconnected stopper for $(T,F)$.

\begin{lemma}\label{lem:rootDis}
Suppose that $\alpha$ is a root or disconnected stopper for $(T,F)$. 
Then, $(T,F)$ has a good normal key $\kappa = (L_\alpha, B)$.
\end{lemma}
\begin{proof}
Let $\gamma_1$ and $\gamma_2$ be the children of $\alpha$ in $T$. Since no descendant 
of $\alpha$ in $T$ is a close stopper for $(T,F)$, $N_{L_\alpha}(v) \le 1$ for every 
descendant $v$ of $\ell_F(L_\alpha)$ in $F$. So, we can construct a fair normal key 
$\kappa_i=(L_{\gamma_i}, B_i)$ of $(T,F)$ for each $i\in\{1,2\}$, as in Lemma~\ref{lem:goUp}. 
Now, by Statement~1 in Lemma~\ref{lem:helpNormal}, we can combine $\kappa_1$ and $\kappa_2$ 
into a good normal key $\kappa=(L_\alpha,B)$ of $(T,F)$. 
\end{proof}


\begin{lemma}\label{lem:stopGoingUp}
Suppose that $\alpha$ is a close stopper for $(T,F)$. Then, 
$(T,F)$ has a good abnormal key $\kappa = (L_\alpha,B,R)$.
\end{lemma}
\begin{proof}
For convenience, let $X=L_\alpha$, $v = \ell_F(X)$, and $u_1$ and $u_2$ be the children of 
$v$ in $F$. Clearly, $N_X(v) = N_{X}(u_1) + N_{X}(u_2)$. So, $N_X(u_1) = N_X(u_2) = 1$. 
For each $i\in\{1,2\}$, let $x_i \in X$ be the leaf endpoint of the unique $X$-path 
starting at $u_i$. Further let $v_{i,0}$, \ldots, $v_{i,k_i}$ be the vertices of 
$x_i \sim_F u_i$, where $v_{i,0} = x_i$, $v_{i,k_i} = u_i$, and $v_{i,j}$ is 
the parent of $v_{i,j-1}$ in $F$ for each $1\le j\le k_i$. 
For each $j\in\{1,\ldots,k_i\}$, let $u_{i,j}$ be the sibling of $v_{i,j-1}$ in $F$. 
For convenience, let $X_{i,j} = X^F(u_{i,j})$. See Figure~\ref{fig:closeStop}. 

\begin{figure}[htb]
\centerline{\includegraphics{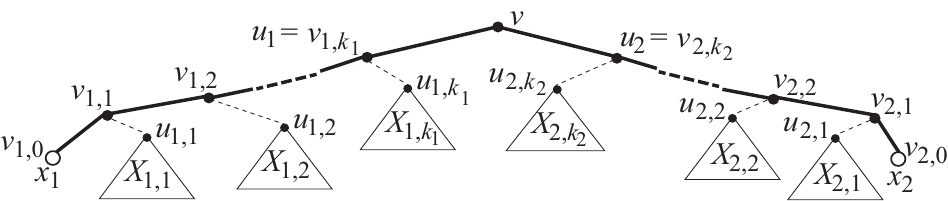}}
\caption{The subtree of $F$ rooted at $v$ in the proof of Lemma~\ref{lem:stopGoingUp}, where 
thin dashed lines show edges in $B$ and each 
triangle shows a dangling subtree whose leaf set is a superset of the set inside the triangle. 
}
\label{fig:closeStop}
\end{figure}

For each $1\le j\le k_i$, if $\ell_F(X_{i,j}) = u_{i,j}$, then $N_X(u_{i,j}) = 0$ 
(because $N_X(v_{i,j}) \le 1$) and hence Lemma~\ref{lem:goUp} guarantees that 
$(T,F)$ has a fair robust key $\kappa_{i,j} = (X_{i,j}, B_{i,j})$; otherwise, 
$N_X(\ell_F(X_{i,j})) \le 1$ and hence Lemma~\ref{lem:goUp} guarantees that 
$(T,F)$ has a fair normal key $\kappa_{i,j} = (X_{i,j}, B_{i,j})$. 
So, $\kappa = (X, B_1\cup B'_1\cup B_2\cup B'_2, R)$ is a key of $(T,F)$, where 
$R$ is the edge set of $x_1 \sim_F x_2$, $B_i = \bigcup_{j=1}^{k_i} B_{i,j}$ 
and $B'_i = \{e_F(u_{i,j})~|~1\le j\le k_i\}$ for each $i\in\{1,2\}$. 

Note that $|\kappa| = k_1 + k_2 + \sum_{i=1}^2\sum_{j=1}^{k_i}|\kappa_{i,j}|$. 
Let $\Gamma_{i,j}$ be an $X_{i,j}$-search tree of $(T,F)$ witnessing $b(\kappa_{i,j})$. 
Obviously, we can use an $X$-search tree $\Gamma$ of $(T,F)$ in which each root-leaf path $P$ 
can be cut into $2(k_1+k_2) + 1$ portions $P_{1,1}$, \ldots, $P_{1,k_1}$, $P_{2,1}$, \ldots, 
$P_{2,k_2}$, $Q_{1,1}$, \ldots, $Q_{1,k_1}$, $Q_{2,1}$, \ldots, $Q_{2,k_2}$, $Q$ 
(appearing in $P$ in this order), where 
\begin{itemize}
\item for each $i\in\{1,2\}$ and each $j\in\{1,\ldots,k_i\}$, 
$P_{i,j}$ corresponds to a root-leaf path in $\Gamma_{i,j}$ and $Q_{i,j}$ is a single edge 
corresponding to picking a cherry $(Y_{i,j}, Z_{i,j})$ with $Y_{i,j}\subseteq X_{i,j}$ and 
$Z_{i,j}\subseteq \bigcup^{j-1}_{h=1} X_{i,h} \cup \{x_i\}$, 

\item $Q$ is a single edge corresponding to picking a cherry $(W_1, W_2)$ with 
$W_i \subseteq \bigcup^{k_i}_{j=1} X_{i,j} \cup \{x_i\}$ for each $i\in\{1,2\}$. 
\end{itemize}
Obviously, the following claims follow from Lemma~\ref{lem:consistent}:

\medskip

{\bf Claim 1.} For each $i\in\{1,2\}$, $Z_{i,1} = \{x_i\}$ and for each $j\in\{2,\ldots,k_i\}$, 
either $\ell_F(Z_{i,j})$ is a vertex of $x_i \sim_F v_{i,j}$ other than $v_{i,j}$ or 
$Z_{i,j}\subseteq X_{i,h}$ for some $h\in\{1,\ldots,j-1\}$. 

\medskip

{\bf Claim 2.} For each $i\in\{1,2\}$ and each $j\in\{1,\ldots,k_i\}$, 
$f_e(\kappa_{i,j},P_{i,j}) \subseteq f_e(\kappa,P)$ and 
$f_c(\kappa_{i,j},P_{i,j}) \subseteq f_c(\kappa,P)$. Moreover, for two pairs
$\{i,j\}$ and $\{i',j'\}$ with $\{i,j\} \ne \{i',j'\}$, 
$f_e(\kappa_{i,j},P_{i,j}) \cap f_e(\kappa_{i',j'},P_{i',j'}) = \emptyset$ and
$f_c(\kappa_{i,j},P_{i,j}) \cap f_c(\kappa_{i',j'},P_{i',j'}) = \emptyset$. 

\medskip

Let $A_e = \bigcup_{i=1}^2\bigcup_{j=1}^{k_i} f_e(\kappa_{i,j},P_{i,j})$ 
and $A_c = \bigcup_{i=1}^2\bigcup_{j=1}^{k_i} f_c(\kappa_{i,j},P_{i,j})$. 
By Claims~1 and~2, $A_e \subseteq f_e(\kappa,P)$, $A_c \subseteq f_c(\kappa,P)$, 
$|A_e| = \sum_{i=1}^2\sum_{j=1}^{k_i} |f_e(\kappa_{i,j},P_{i,j})|$, and 
$|A_c| = \sum_{i=1}^2\sum_{j=1}^{k_i} |f_c(\kappa_{i,j},P_{i,j})|$. 
For each $i\in\{1,2\}$ and each $j\in\{1,\ldots,k_i\}$, $e_F(u_{i,j}) \not\in B_{i,j}$ 
(and hence $e_F(u_{i,j}) \not\in A_e$) because $D_F(v_{i,j}, \ell_F(X_{i,j})) \ne 
\emptyset$ or $\kappa_{i,j}$ is robust. 

Consider an $i\in\{1,2\}$ and a $j\in\{1,\ldots,k_i\}$. Further consider the node 
$\varpi$ of $\Gamma$ at which $P$ picks the cherry $(Y_{i,j}, Z_{i,j})$. 
For convenience, let $T'=T\ominus C_T$ and $F'=F\ominus C(\varpi)$, where $C_T$ is 
the cut of $T$ forced by $C(\varpi)$. By Statement~1 in Lemma~\ref{lem:consistent}, 
$\ell_{T'}(Y_{i,j})$ agrees with a vertex $w_{i,j}$ of $F'$ 
that is a descendant of $\ell_F(X_{i,j})$ in $F$, while $\ell_{T'}(Z_{i,j})$ agrees with 
a vertex $w'_{i,j}$ of $F'$ that is a descendant of $v_{i,j-1}$ in $F$. Moreover, 
by Claim~1, either $w'_{i,j}$ is a vertex of $x_i \sim_F v_{i,j-1}$ or 
$w'_{i,j}$ is a descendant of $\ell_F(X_{i,h})$ in $F$ for some $h\in\{1,\ldots,j-1\}$. 
Let $e_{i,j}$ (respectively, $e'_{i,j}$) be the edge of $F$ corresponding to $e_{F'}(w_{i,j})$ 
(respectively, $e_{F'}(w'_{i,j})$). 

\medskip

{\bf Claim 3.} For each $i\in\{1,2\}$ and each $j\in\{1,\ldots,k_i\}$, 
(1)~$f_c(\kappa,P)\setminus A_c$ contains a component rooted at $u_{i,j}$, 
(2)~$e'_{i,j} \in C(P) \cap R$, (3)~$e'_{i,j} \in C(P)\setminus R$ and 
either $e'_{i,j} \in f_e(\kappa,P) \setminus A_e$ or $f_c(\kappa,P)\setminus A_c$ 
contains a connected component rooted at the head of $e'_{i,j}$, or 
(4)~$e_{i,j} \in C(P)\setminus R$ and either $e_{i,j} \in f_e(\kappa,P)\setminus A_e$ or 
$f_c(\kappa,P)\setminus A_c$ contains a connected component rooted at the head of $e_{i,j}$. 

\medskip

To prove Claim~3, consider the three possible ways in which $P$ picks the cherry 
$(Y_{i,j}, Z_{i,j})$. (1)~holds when $P$ creates $Y_{i,j}\cup Z_{i,j}$, because of 
Statement~3 in Lemma~\ref{lem:mergeHelp}. (2)~holds when $P$ isolates $Z_{i,j}$ and 
$w'_{i,j}$ is a vertex of $x_i \sim_F v_{i,j-1}$. 
(3)~holds when $P$ isolates $Z_{i,j}$ and $w'_{i,j}$ is a descendant of $\ell_F(X_{i,h})$ 
in $F$ for some $h\in\{1,\ldots,j-1\}$, because of Lemma~\ref{lem:cutHelp}.
Similarly, (4)~holds when $P$ isolates $Y_{i,j}$. 

By Claim~3, each pair $(i,j)$ with $i\in\{1,2\}$ and $j\in\{1,\ldots,k_i\}$ contributes 
at least one element to $(f_e(\kappa,P)\setminus A_e) \cup (f_c(\kappa,P) \setminus A_c)$. 

\medskip

{\bf Claim 4.} Two different pairs $(i,j)$ and $(i',j')$ contribute different elements 
to $(f_e(\kappa,P)\setminus A_e) \cup (f_c(\kappa,P) \setminus A_c)$. 

\medskip

Claim~4 is clearly true when $i \ne i'$. So, assume that $i = i'$ and $j < j'$. 
If (1)~in Claim~3 holds for $(i,j)$, then after $P$ picks $(Y_{i,j}, Z_{i,j})$, 
$e_F(u_{i,j})$ cannot be cut by $P$ when picking $(Y_{i,j'},Z_{i,j'})$. 
Moreover, if (2) or (3)~in Claim~3 holds for $(i,j)$, then after $P$ picks 
$(Y_{i,j}, Z_{i,j})$, $e'_{i,j}$ has already been cut by $P$ and hence cannot be cut 
by $P$ again when picking $(Y_{i,j'},Z_{i,j'})$. Similarly, if (4)~in Claim~3 holds for 
$(i,j)$, then after $P$ picks $(Y_{i,j}, Z_{i,j})$, $e_{i,j}$ has already been cut by $P$ 
and hence cannot be cut by $P$ again when picking $(Y_{i,j'},Z_{i,j'})$. Thus, in any case, 
$(i,j)$ and $(i',j')$ contribute different elements to 
$(f_e(\kappa,P)\setminus A_e) \cup (f_c(\kappa,P) \setminus A_c)$, i.e., Claim~4 holds. 

By Claims 3 and 4, the $k_1+k_2$ pairs $(i,j)$ with $i\in\{1,2\}$ and $j\in\{1,\ldots,k_i\}$ 
contributes at least a total of $k_1+k_2$ elements to $(f_e(\kappa,P) \setminus A_e) \cup 
(f_c(\kappa,P) \setminus A_c)$. Thus, 
$b(\kappa,P) \ge k_1 + k_2 + \sum_{i=1}^2\sum_{j=1}^{k_i} b(\kappa_{i,j},P_{i,j})
\ge k_1 + k_2 + \sum_{i=1}^2\sum_{j=1}^{k_i} b(\kappa_{i,j})$. Recall that 
$|\kappa| = k_1 + k_2 + \sum_{i=1}^2\sum_{j=1}^{k_i}|\kappa_{i,j}|$. Since 
$|\kappa_{i,j}| \le 2b(\kappa_{i,j}) + 1$, $|\kappa| \le 2(k_1+k_2) + 
2\sum_{i=1}^2\sum_{j=1}^{k_i} b(\kappa_{i,j}) \le 2b(\kappa,P)$. 
Because $P$ is an arbitrary root-leaf path of $\Gamma_X$, $|\kappa| \le 2b(\kappa)$.
\end{proof}


\begin{lemma}\label{lem:stopGoingUp2}
Suppose that $\alpha$ is a semi-close stopper for $(T,F)$. Then, 
$(T,F)$ has a good abnormal key $\hat{\kappa} = (L_\alpha, \hat{B}, R)$.
\end{lemma}
\begin{proof}
Let $x_1$ and $x_2$ be as in the definition of a close stopper for $(T,F)$. 
Let $A$ be the set of $L_\alpha$-exclusive edges in $D_F(x_1,x_2)$. 
By Lemma~\ref{lem:stopGoingUp}, the lemma holds if $A=\emptyset$. So, we assume that 
$A \ne \emptyset$. Clearly, we can construct a good abnormal key $\kappa=(L_\alpha,B,R)$ 
of $(T,F\ominus A)$ as in the proof of Lemma~\ref{lem:stopGoingUp}. Thus, we hereafter 
inherit the notations in that proof. Recall that $R$ is the edge set of the path between 
$x_1$ and $x_2$ in $F\ominus A$. Further note that the edges in $B$ also appear in $F$. 

We set $\hat{B} = B \cup A$. Obviously, $\hat{\kappa}=(X,\hat{B},\hat{R})$ is 
an abnormal key of $(T,F)$, where $\hat{R}$ is the edge set of $x_1\sim_F x_2$. 
Since $|\hat{\kappa}| = |A| + k_1 + k_2 + \sum_{i=1}^2\sum_{j=1}^{k_i}|\kappa_{i,j}|$ 
and $|A| \le 2$, it remains to show that $b(\hat{\kappa},P) \ge 1 + k_1 + k_2 + 
\sum_{i=1}^2\sum_{j=1}^{k_i} b(\kappa_{i,j},P_{i,j})$. 
The proof is a simple extension of that of Lemma~\ref{lem:stopGoingUp}. Indeed, the 
proof of Lemma~\ref{lem:stopGoingUp} (as it is) already shows that $b(\hat{\kappa},P) 
\ge k_1 + k_2 + \sum_{i=1}^2\sum_{j=1}^{k_i} b(\kappa_{i,j},P_{i,j})$. We want to show 
that we can always increase this lower bound by~1. To this end, we take a closer look 
at (1) in Claim~3 and its proof by distinguishing two cases as follows. 

\medskip
{\em Case 1:} $P$ picks $(Y_{i,j},Z_{i,j})$ by creating $Y_{i,j} \cup Z_{i,j}$, and 
$w_{i,j} \sim_F w'_{i,j}$ contains the tail of some edge $\hat{e} \in A$. In this case, 
not only $f_c(\kappa,P)\setminus A_c$ contains a component rooted at $u_{i,j}$ but also 
$\hat{e} \in C(P) \cap \hat{B}$. 

\medskip
{\em Case 2:} $P$ picks $(Y_{i,j},Z_{i,j})$ by creating $Y_{i,j} \cup Z_{i,j}$, and 
$w'_{i,j}$ is a descendant of $\ell_F(X_{i,h})$ in $F$ for some $h\in\{1,\ldots,j-1\}$. 
In this case, not only $f_c(\kappa,P)\setminus A_c$ contains a component rooted at $u_{i,j}$ 
but also $C(P) \cap \hat{R}$ contains the edge entering the child of $v_{i,h}$ that is 
not an ancestor of $\ell_F(X_{i,h})$. 

\medskip
If Case~1 or~2 happens for some $i\in\{1,2\}$ and $j\in\{1,\ldots,k_i\}$, then we have 
increased the lower bound $k_1 + k_2 + \sum_{i=1}^2\sum_{j=1}^{k_i} b(\kappa_{i,j},P_{i,j})$ 
on $b(\hat{\kappa},P)$ by~1 and hence are done. So, assume that neither Case~1 nor~2 ever 
occurs. Let $\tilde{e}$ be the edge in $A$ whose tail is the furthest from $v$ in $F$ 
among the tails of the edges in $A$. Then, there are an $i\in\{1,2\}$ and 
a $j \in \{0,\ldots,k_i\}$ such that the tail of $\tilde{e}$ is an inner vertex of 
$v_{i,j} \sim_F v_{i,j+1}$, where we define $v_{i,k_i+1} = v$. We may assume $i=1$. 

Now, consider the node $\varrho$ of $\Gamma$ at which $P$ picks the final cherry $(W_1,W_2)$. 
For convenience, let $T'' = T \ominus C'_T$ and $F'' = F \ominus C(\varrho)$, where $C'_T$ is 
the cut of $T$ forced by $C(\varrho)$. 
In the proof of Lemma~\ref{lem:stopGoingUp}, we ignored the contribution of $(W_1,W_2)$ 
to the lower bound $k_1 + k_2 + \sum_{i=1}^2\sum_{j=1}^{k_i} b(\kappa_{i,j},P_{i,j})$ 
on $b(\hat{\kappa},P)$. So, we here want to count the contribution. For each $i\in\{1,2\}$, 
let $w_i$ be the vertex in $F''$ agreeing with $\ell_{T''}(W_i)$ in $T''$. Since 
neither Case~1 nor~2 occurs, either $w_1$ is a descendant of $v_{1,j}$ in $F''$, or $j < k_1$ 
and $w_1$ is a descendant of $\ell_F(X_{1,h})$ in $F$ for some $h\in\{j+1,\ldots,k_1\}$. 
In the former case, $\tilde{e} \in D_F(w_1,w_2)$, implying that $\tilde{e} \in 
f_e(\hat{\kappa},P)$ if $P$ creates $W_1 \cup W_2$. In the latter case, $D_F(w_1,w_2)$ 
contains the edge $\ddot{e}$ entering the child of $v_{1,h}$ that is not an ancestor of 
$u_{1,h}$, implying that $\ddot{e} \in C(P) \cap \hat{R}$ if $P$ creates $W_1 \cup W_2$. 
On the other hand, if $P$ isolates $W_i$ for some $i\in\{1,2\}$, then 
either $f_e(\hat{\kappa},P)$ contains the edge $e''_i$ of $F$ corresponding to $e_{F''}(w_i)$ 
or $f_c(\hat{\kappa},P)$ contains a component rooted at the head of $e''_i$. Therefore, 
in any case, we can increase the lower bound 
$k_1 + k_2 + \sum_{i=1}^2\sum_{j=1}^{k_i} b(\kappa_{i,j},P_{i,j})$ on $b(\hat{\kappa},P)$ by~1. 
\end{proof}

\subsection{The Hard Case}\label{sec:overlap}
A vertex $\alpha$ of $T$ is an {\em overlapping stopper} for $(T,F)$ if $\ell_F(L_\alpha)$ is 
defined, no descendant of $\alpha$ in $T$ is a semi-close, root, or disconnected stopper for 
$(T,F)$, and both children $\lambda_1$ and $\lambda_2$ of $\alpha$ in $T$ are consistent with 
$F$ but $\alpha$ is not. For example, if $T$ and $F$ are as in Figure~\ref{fig:def}, then 
both $\alpha$ and $\beta$ are overlapping stoppers for $(T,F)$. 

\begin{lemma}\label{lem:exist}
There always exists a semi-close, root, disconnected, or overlapping stopper for $(T,F)$. 
\end{lemma}
\begin{proof}
By assumption, the root in $T$ agrees with no vertex in $F$. So, $T$ has a vertex $\alpha$ 
such that $\alpha$ is not consistent with $F$ but both children of $\alpha$ are. If some proper 
descendant of $\alpha$ in $T$ is a semi-close stopper for $(T,F)$, we are done. So, assume that 
$\alpha$ has no such proper descendant in $T$. Now, if some proper descendant of $\alpha$ in $T$ 
is a root stopper for $(T,F)$, we are done; otherwise, $\alpha$ must be a disconnected or 
overlapping stopper for $(T,F)$. 
\end{proof}

So, it remains to show how to compute a good cut for $(T,F)$ when an overlapping stopper 
for $(T,F)$ exists. The next lemma will be useful later. 

\begin{lemma}\label{lem:2dis}
Suppose that $\alpha_1$ through $\alpha_k$ are two or more pairwise incomparable vertices of $T$ 
satisfying the following conditions: 
\begin{itemize}
\item For every $i \in \{1,\ldots,k\}$, $\alpha_i$ is consistent with $F$ and 
	$N_{L_{\alpha_i}}(v) \le 1$ for each descendant $v$ of $\ell_F(L_{\alpha_i})$ in $F$. 
\item $\bigcup^k_{i=1} L_{\alpha_i} = L_\alpha$, where $\alpha=\ell_T(\alpha_1,\ldots,\alpha_k)$. 
\item $\ell_F(L_{\alpha_i} \cup L_{\alpha_j})$ is undefined for every pair $(i,j)$ with $1\le i < j\le k$. 
\item Either $\ell_F(L_{\alpha_i})$ is a root of $F$ for every $i\in\{1,\ldots,k\}$, or 
neither $\ell_F(L_{\alpha_1})$ nor $\ell_F(L_{\alpha_2})$ is a root in $F$.
\end{itemize}
Then, $(T,F)$ has a good cut $C$ with $\{e_F(x)~|~x\in L_\alpha \mbox{ is not a root of } F\} \subseteq C$. 
\end{lemma}
\begin{proof}
Let ${\cal I}_1$ be the set of all $i\in\{1,\ldots,k\}$ such that $\ell_F(L_{\alpha_i})$ is a root 
in $F$ and $|L_{\alpha_i}| \ge 2$. Further let ${\cal I}_2$ be the set of all $i \in \{1,\ldots,k\}$ 
such that $\ell_F(L_{\alpha_i})$ is not a root in $F$. Without loss of generality, we may assume that 
${\cal I}_2 = \{1, 2, \ldots, m\}$. For each $i \in {\cal I}_1$, we construct 
a good normal key $\kappa_i = (L_{\alpha_i},B_i)$ of $(T,F)$ as in Lemma~\ref{lem:rootDis}. 
Let $C_1 = \bigcup_{i\in{\cal I}_1}B_i$ and $X_1 = \bigcup_{i\in{\cal I}_1}L_{\alpha_i}$. 
By Lemma~\ref{lem:rootDis}, $C_1$ is a good cut for $(T,F)$ with $\{e_F(x)~|~x\in X_1\} \subseteq C_1$. 
So, we are done if $m = 0$. Thus, we may assume that $m \ge 2$. Let $(T',F')$ be the sub-TF of $(T,F)$ 
induced by $C_1$. Obviously, $F'$ is a sub-forest of $F$. Moreover, the conditions in the lemma 
still hold after replacing $T$, $F$, and $k$ with $T'$, $F'$, and $m$, respectively. 
For each $i \in {\cal I}_2$, we use Lemma~\ref{lem:goUp} to construct a fair normal key 
$\kappa_i = (L_{\alpha_i},B_i)$. Let $C_2 = \bigcup_{i\in{\cal I}_2}B_i$ and 
$X_2 = \bigcup_{i\in{\cal I}_2}L_{\alpha_i}$. Now, to show that $C = C_1 \cup C_2$ is desired, 
it suffices to show that $\kappa = (X_2, C_2)$ is a good normal key for $(T',F')$. 

For each $i \in {\cal I}_2$, let $\Gamma_i$ be an $L_{\alpha_i}$-search tree of $(T',F')$ witnessing 
$b(\kappa_i)$. We use an $X_2$-search tree $\Gamma$ in which each root-leaf path $P$ can be 
cut into $m+1$ paths $P_1$, \ldots, $P_{m}$, $Q$, where each $P_i$ with $i \in {\cal I}_2$ 
corresponds to a root-leaf path in $\Gamma_i$. For each $i\in{\cal I}_2$, let $Z_i$ be the set of 
$L_{\alpha_i}$-descendants of $\ell_{T'}(L_{\alpha_i})$ in $T' - C_{T',i}$, where $C_{T',i}$ is 
the cut of $T'$ forced by $C(P_i)$. Further let $e_i$ be the edge of $F'$ corresponding to 
$e_{F''}(Z_i)$, where $F'' = F' \ominus C(P_i)$. 

The first edge of $Q$ corresponds to picking a cherry $(Z_i,Z_j)$ for some $1\le i<j\le m$ such 
that $\ell_{T'}(L_{\alpha_i})$ and $\ell_{T'}(L_{\alpha_j})$ are siblings in $T'$. Picking $(Z_i,Z_j)$ 
is done by selecting $h\in\{i,j\}$ and adding $e_h$ to $C(P)$; as the result, 
Lemma~\ref{lem:cutHelp} ensures that either $e_h$ becomes an edge in $f_e(\kappa,P)$ or 
$f_c(\kappa,P)$ contains a component rooted at the head of $e_h$. So, in any case, after 
picking $(Z_i,Z_j)$, $Z_{h'}$ will be one part of a cherry for $P$ to pick later, where $h'$ is 
the integer in $\{i,j\}$ such that $e_{h'}$ was not added to $C(P)$ when picking 
$(Z_i,Z_j)$; consequently, the next cherry for $P$ to pick is still of the form $(Z_{i'}, Z_{j'})$ 
for some $1\le i'<j'\le m$. Thus, each edge of $Q$ corresponds to picking a cherry of this form. 
Moreover, after $P$ finishes picking cherries, there is at most one $g \in {\cal I}_2$ such that 
$e_g \not\in C(P)$. Now, $|\kappa| = \sum_{i\in{\cal I}_2}|\kappa_i| \le 2\sum_{i\in{\cal 
I}_2}b(\kappa_i,P_i) + m$ and $b(\kappa,P) \ge \sum_{i\in{\cal I}_2} b(\kappa_i,P_i) + (m - 1)$. 
Since $m\ge 2$, $|\kappa| \le 2b(\kappa,P)$. 
\end{proof}

Hereafter, fix an overlapping stopper $\alpha$ for $(T,F)$. 
We show how to use $\alpha$ to compute a good cut for $(T,F)$ below. 
Let $X = L_\alpha$, and $X_1$ (respectively, $X_2$) be the set of leaf descendants of 
the left (respectively, right) child of $\alpha$ in $T$. 

Two leaves $a$ and $b$ are {\em far apart} in $F$ if $|D_F(a,b)| \ge 3$. 
Since no descendant of $\alpha$ in $T$ is a semi-close stopper for $(T,F)$, 
every two leaves in $L_\alpha$ are far apart in $F$. 
The next lemma strengthens Lemma~\ref{lem:goUp}. 

\begin{lemma}\label{lem:useful}
Let $\beta$ be a non-leaf proper descendant of $\alpha$ in $T$. 
Then, $(T,F)$ has a fair robust key $\kappa=(L_\beta,B)$ such that 
if no child of $\ell_F(L_\beta)$ in $F$ is a leaf, $\kappa$ is super-robust. 
\end{lemma}
\begin{proof} For convenience, let $Y=L_\beta$. 
Let $I$ be the set of $Y$-bifurcate vertices in $F$. 
We claim that for every $v\in I$ with $Y$-children $v_1$ and $v_2$ in $F$, $(T,F)$ has 
a fair normal key $\kappa_v = (Y^F(v), B_v)$ such that (1)~$\kappa_v$ is robust, 
and (2)~$\kappa_v$ is super-robust if neither $v_1$ nor $v_2$ is both a leaf and a child 
of $v$ in $F$. 
We prove the claim by induction on $|Y^F(v)|$. By Lemma~\ref{lem:trivial2}, the claim is 
true if $|Y^F(v)| = 2$. So, assume that $|Y^F(v)| \ge 3$. We choose an edge $\tilde{e}$ 
via the robust way of combining $\kappa_{v_1}$ and $\kappa_{v_2}$. We set $B_v = B_{v_1} 
\cup B_{v_2} \cup \{\tilde{e}\}$ and $\kappa_v = (Y^F(v),B_v)$. By Statement~2 
in Lemma~\ref{lem:helpNormal}, $|\kappa_v| \le 2b(\kappa_v) + 1$. 

\medskip
{\em Case 1:} For some $i\in\{1,2\}$, $v_i$ is both a leaf and a child of $v$ in $F$. 
We may assume $i=1$. Then, $v_1\in Y$ by the above crucial point. Moreover, since every 
two leaves in $Y$ are far apart in $F$, either neither child of $v_2$ in $F$ is a leaf, 
or $|D_F(v,v_2)|\ge 2$ and some child of $v_2$ in $F$ is a leaf. In the former case, 
$\kappa_{v_2}$ is super-robust by the inductive hypothesis, and hence $\kappa_v$ is robust by 
the choice of $\tilde{e}$. In the latter case, $\kappa_v$ is robust by the choice of $\tilde{e}$. 

\medskip
{\em Case 2:} Case~1 does not occur. Then, for each $i\in\{1,2\}$, $D_F(v,v_i)\ne\emptyset$, 
or $\kappa_{v_i}$ is robust by the inductive hypothesis. So, if we can claim that for some 
$i\in\{1,2\}$, $|D_F(v_i,v)| \ge 2$, $\kappa_{v_i}$ is super-robust, or $D_F(v,v_i)\ne\emptyset$ 
and $\kappa_{v_i}$ is robust, then $\kappa_v$ is super-robust by the choice of $\tilde{e}$. 
For a contradiction, assume that the claim is false. Then, for each $i\in\{1,2\}$, 
either (1)~$|D_F(v_i,v)| = 1$ and $\kappa_{v_i}$ is not robust, or 
(2)~$D_F(v,v_i)=\emptyset$ and $\kappa_{v_i}$ is not super-robust. 
Thus, by the inductive hypothesis, we know that for each $i\in\{1,2\}$, 
either (1)~$|D_F(v_i,v)| = 1$ and $v_i$ is a leaf, or 
(2)~$D_F(v,v_i)=\emptyset$ and some child of $v_i$ in $F$ is a leaf. However, 
in any case, we found two leaves in $Y$ that are not far apart in $F$, a contradiction. 
\end{proof}

For an $i\in\{1,2\}$, a vertex $\hat{u}$ of $F$ is an {\em $X_i$-port} if 
$\hat{u}$ is $X_i$-inclusive but $X_{3-i}$-exclusive in $F$, 
$\hat{u}$ is a non-leaf of $F$, and 
$e_F(\hat{u}) \in D_F(x_1,x_2)$ 
for some vertices $x_1$ and $x_2$ in $X_{3-i}$. 


\begin{lemma}\label{lem:port}
Suppose that for some $i\in\{1,2\}$, there is an $X_i$-port $\hat{u}$ in $F$. 
Then, $(T,F)$ has a good normal key. 
\end{lemma}
\begin{proof}
We may assume $i=2$. We use Lemma~\ref{lem:helpNormal} to construct a fair normal key 
$\kappa_1=(X_1,B_1)$ of $(T,F)$ with $e_F(\hat{u}) \in B_1$. Let $X'_2 = X_2^F(\hat{u})$. 
If $|X'_2|\ge 2$, we use Lemma~\ref{lem:useful} to construct a fair robust key 
$\kappa_2=(X'_2,B_2)$ of $(T,F)$; otherwise, $X'_2$ consists of a single 
$x_2\in X_2$ and $\kappa_2=(X'_2,B_2)$ with $B_2=\{e_F(x_2)\}$. 
Obviously, no matter what $|X_2|$ is, $e_F(\hat{u}) \not\in B_2$. 
Moreover, $\kappa = (X_1\cup X'_2, B)$ with $B=B_1\cup B_2$ is a normal key 
of $(T,F)$. We next show that $|\kappa| \le 2b(\kappa)$. 

Clearly, $|\kappa| = |\kappa_1| + |\kappa_2| \le 2(b(\kappa_1) + b(\kappa_2) + 1)$. 
So, it remains to show that $b_X(\kappa) \ge b(\kappa_1) + b(\kappa_2) + 1$. 
To this end, we use an $X$-search tree of $(T,F)$ in which each root-leaf path 
$P$ can be cut into four portions $P_1$, $P_2$, $Q_1$, $Q_2$, where $P_1$ is a root-leaf path 
in an $X_1$-search tree of $(T,F)$ witnessing $b(\kappa_1)$, $P_2$ is a root-leaf path in 
an $X'_2$-search tree of $(T,F)$ witnessing $b(\kappa_2)$, each edge of $Q_1$ corresponds to 
picking a cherry $(Y,Z)$ with $Y\subseteq X_2$ and $Z\subseteq X_2$, and $Q_2$ consists of 
a single edge corresponding to picking a cherry $(Y_1,Y_2)$ with $Y_1\subseteq X_1$ and 
$Y_2 \subseteq X_2$. By Lemma~\ref{lem:consistent}, $f_e(\kappa_1,P_1)\cap f_e(\kappa_2,P_2) = 
\emptyset$, $f_c(\kappa_1,P_1)\cap f_c(\kappa_2,P_2) = \emptyset$, and $f_c(\kappa_2,P_2) 
\subseteq f_c(\kappa,P)$. Moreover, since $C(P_1)\cup C(P_2) \subseteq C(P)$, $f_e(\kappa_1,P_1) 
\cup f_e(\kappa_2,P_2) \subseteq f_e(\kappa,P)$. For the same reason, 
for each $K \in f_c(\kappa_1,P_1)$, $K$ either remains in $f_c(\kappa,P)$ or is split into 
two or more components in $f_c(\kappa,P)$. Thus, $|f_c(\kappa_1,P_1)| + |f_c(\kappa_2,P_2)| 
\le |f_c(\kappa,P)|$. Therefore, $b(\kappa,P) \ge b(\kappa_1,P_1) + b(\kappa_2,P_2)$.
We claim that $b(\kappa,P) \ge b(\kappa_1,P_1) + b(\kappa_2,P_2) + 1$. To see this claim, 
first note that no component in $f_c(\kappa_1,P_1)$ is rooted at $\hat{u}$ because 
each $x \in X'_2$ is a descendant of $\hat{u}$ in $F - C(P_1)$. Similarly, no component in 
$f_c(\kappa_2,P_2)$ is rooted at $\hat{u}$ because $e_F(\hat{u}) \not\in B_2\cup C(P_2)$. 
We next distinguish two cases as follows. 

\medskip
{\em Case 1:} $\hat{u}$ has no labeled descendant in $F - (B_2 \cup C(P_2))$. In this case, 
since $e_F(\hat{u}) \in B_1$, $f_c(\kappa,P)$ contains a component rooted at $\hat{u}$. 
So, $b(\kappa,P) \ge b(\kappa_1,P_1) + b(\kappa_2,P_2) + 1$. 

\medskip
{\em Case 2:} $\hat{u}$ has a labeled descendant in $F - (B_2 \cup C(P_2))$. In this case, 
consider the set $Y_1$ of all $x \in X_1$ such that $x$ is a descendant of $\ell_F(X_1)$ 
in $F - C(P_1)$, and the set $Y_2$ of all $x \in X'_2$ such that $x$ is a descendant of 
$\ell_F(X'_2)$ in $F - C(P_2)$. By Lemma~\ref{lem:consistent}, no component in 
$f_c(\kappa_1,P_1)$ is rooted at an ancestor of $\ell_F(Y_1)$ in $F$ and no component in 
$f_c(\kappa_2,P_2)$ is rooted at a vertex that is both an ancestor of $\ell_F(Y_2)$ in $F$ 
and a descendant of $\hat{u}$ in $F$. So, if for some $i\in\{1,2\}$, $F - C(P)$ has a component 
whose leaf set is $Y_i$, then we are done because by Lemma~\ref{lem:cutHelp}, either $Y_i$ 
consists of a single vertex of $X_i$ and $e_F(\ell_F(Y_i)) \in f_e(\kappa, P)\setminus 
f_e(\kappa_i,P_i)$, or $f_c(\kappa,P) \setminus f_c(\kappa_i,P_i)$ contains a component rooted 
at a vertex that is an ancestor of $\ell_F(Y_i)$ in $F$ (and a descendant of $\hat{u}$ in $F$ 
if $i=2$). So, we can assume that such $i$ does not exist. Then, $Q_2$ must pick the cherry 
$(Y_1,Y_2)$ in the third way and hence $\hat{u}$ is a root of some component in $f_c(\kappa,P)$. 
Therefore, $b(\kappa, P') \ge b(\kappa_1, P_1) + b(\kappa_2, P_2) + 1$. 
\end{proof}

By Lemma~\ref{lem:port}, we may hereafter assume that 
there is neither $X_1$-port nor $X_2$-port in $F$. 

\begin{lemma}\label{lem:dangle}
Suppose that for some $h\in\{1,2\}$, $\ell_F(X_h)$ is a descendant of $\ell_F(X_{3-h})$ and 
is $X_{3-h}$-exclusive in $F$. Then, there is a good cut in $F$. 
\end{lemma}
\begin{proof} 
We may assume $h=2$. Then, $\ell_F(X_1)$ is an ancestor of $\ell_F(X_2)$ in $F$. Further let 
$v_1$, \ldots, $v_k$ be those $X_1$-bifurcate vertices in $F$ that are also $X_2$-inclusive 
in $F$, where $v_i$ is the $X_1$-parent of $v_{i-1}$ in $F$ for each $i\in\{2,\ldots,k\}$. 
Let $v'_0$ and $v'_1$ be the $X_1$-children of $v_1$ in $F$, where $\ell_F(X_2)$ is 
a descendant of the head of some $\hat{e} \in D_F(v'_0,v_1)$ in $F$. 
For each $i\in\{2,\ldots,k\}$, let $v'_{i}$ be the other $X_1$-child of $v_i$ in $F$. 
Since there is no $X_2$-port in $F$, $X_2$ consists of a single $x_2$ and $\hat{e} = e_F(x_2)$. 
Let $u$ be the tail of $\hat{e}$. See Figure~\ref{fig:dangleStop}. 

\begin{figure}[htb]
\centerline{\includegraphics{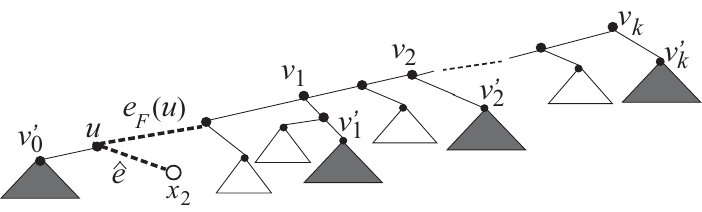}}
\caption{The subtree of $F$ rooted at $\ell_F(X_1)$ in the proof of Lemma~\ref{lem:dangle},
where filled triangles show dangling subtrees containing at least one leaf in $X_1$, unfilled 
triangles show dangling subtrees containing no leaf in $X_1$, and bold dashed lines show edges 
in $B$. }
\label{fig:dangleStop}
\end{figure}

It is possible that $v'_0 \in X_1$ and $v'_0$ and $x_2$ are siblings in $F$. 
Suppose that this indeed happens. 
Then, we can use Lemma~\ref{lem:goUp} to construct a fair normal key $\kappa = (X_1, B)$ of 
$(T,F)$ with $\hat{e} \in B$. Obviously, $F \ominus B$ has $|F| + |B|$ components. Moreover, 
since $|\kappa| \le 2b(\kappa) + 1$, $d(T,F \ominus B) \le d(T,F) - \frac{|B|-1}{2}$. 
Let $\tilde{F}$ be obtained from $F\ominus B$ by adding back the parent of $v'_0$ and $x_2$ 
together with $e_F(v'_0)$ and $\hat{e}$. Clearly, $d(T,\tilde{F}) = 
d(T,F\ominus B)$ but $|\tilde{F}| = |F\ominus B|-1$. Hence, $|\tilde{F}| - |F| \le 2(d(T,F) - 
d(T,\tilde{F}))$. Thus, $(B\setminus\{\hat{e},e_F(v'_0)\})\cup \{e_F(u)\}$ is a good cut of $F$. 
So, we hereafter assume that $v'_0 \not\in X_1$ or $D_F(v'_0,u) \ne\emptyset$. 

We use Lemma~\ref{lem:useful} to construct a fair robust key $\kappa = (X_1, B)$ of $(T,F)$. 
For each $i\in\{0,1,\ldots,k\}$, consider $\kappa_i = (X_1^F(v'_i), B_i)$, where $B_i$ is the set 
of all $e \in B$ such that the head of $e$ is a descendant of $v'_i$ in $F$. We may assume that 
$\hat{e} \in B$, because (1)~we choose edges via the robust way of combining keys in the construction 
of $\kappa$, and (2)~$D_F(v'_0,u )\ne \emptyset$ or $\kappa_i$ is a robust key of $(T,F)$. 
We claim that $\tau = (X_1, B \cup \{e_F(u)\})$ is a normal key of $(T,F)$. Since $v'_0 \not\in X_1$ 
or $D_F(v'_0,u) \ne\emptyset$, Lemma~\ref{lem:useful} guarantees that $u$ has a labeled descendant 
in $F - (B\cup \{e_F(u)\})$. Hence, by Lemma~\ref{lem:useful}, the claim is true 
if $v'_1 \not\in X_1$, $D_F(v'_1,u) \ne \emptyset$, or $X_1$ contains neither $v'_2$ nor a child 
of $v'_2$ in $F$. So, assume that $v'_1 \in X_1$, $D_F(v'_1,u) = \emptyset$, and $X_1$ contains 
$v'_2$ or a child of $v'_2$ in $F$. If $v'_2 \in X_1$, then $D_F(v_1,v'_2) \ge 2$ (because $v'_1$ 
and $v'_2$ are far apart in $F$) and $B$ contains at most one edge in $D_F(v_1,v'_2)$. On the other 
hand, if $X_1$ contains a child of $v'_2$ in $F$, then $D_F(v_1,v'_2) \ge 1$ (because every two 
leaves in $X_1$ are far apart in $F$), and in turn Lemma~\ref{lem:useful} guarantees that 
$B_{2}$ contains one edge in $D_F(v_1,v'_2)$. Therefore, in any case, the component in 
$F - (B \cup \{e_F(u)\})$ containing $v_1$ also contains at least one labeled leaf of $F$. 
Consequently, the claim always holds. 

In summary, we have constructed a fair normal key $\kappa = (X_1, B)$ of $(T,F)$ such that 
$\hat{e} \in B$ and $\tau = (X_1, \hat{B})$ with $\hat{B}=B\cup \{e_F(u)\}$ is a normal key 
of $(T,F)$. Let $\Gamma_1$ be an $X_1$-search tree $\Gamma_1$ of $(T,F)$ witnessing $b(\kappa)$. 
Further let $\Gamma$ be an $X$-search tree of $(T,F)$ in which each root-leaf path $P$ can be 
transformed into a root-leaf path $P_1$ of $\Gamma_1$ by deleting the last edge. Let $P$ be an 
arbitrary root-leaf path in $\Gamma$, and $P_1$ be obtained from $P$ by deleting the last edge. 
Since $\hat{B}$ is a superset of $B$, $f_e(\kappa,P_1) \subseteq f_e(\tau, P_1)$ and 
$|f_c(\tau,P_1)| \ge |f_c(\kappa, P_1)|$. Moreover, if $u$ has no labeled descendant in 
$F - (B \cup C(P_1))$, then $|f_c(\tau,P_1)| > |f_c(\kappa,P_1)|$ and in turn 
$b(\tau,P_1) \ge b(\kappa,P_1) + 1$, implying that $\tau$ is a good key of $(T,F)$ 
because $|\tau| = |\kappa|+1 \le 2b(\kappa) + 2$. Thus, we may assume that $u$ has a labeled 
descendant in $F - (B \cup C(P_1))$. By this assumption, $P_1$ never picks a cherry $(Y,Z)$ with 
$Y \subseteq X^F_1(u)$ and $Z \cap X^F_1(u) = \emptyset$ in the third way. 
Hence, one of the following cases occurs: 

\medskip
{\em Case 1:} The set of leaf descendants of $\ell_F(X_1)$ in $F - C(P_1)$ is a 
$Y \subseteq X^F_1(v'_0)$. In this case, the last cherry for $P$ to pick is $(Y, \{x_2\})$ 
and $P$ isolates $Y$ or $\{x_2\}$. If $P$ isolates $\{x_2\}$, then $\hat{e} \in f_e(\tau,P_1) 
\setminus f_e(\kappa,P_1)$; otherwise, by Lemma~\ref{lem:cutHelp}, $f_e(\tau,P) \setminus 
f_e(\kappa,P_1) \ne \emptyset$ or $f_c(\tau,P) \setminus f_c(\kappa,P_1) \ne \emptyset$. 
So, in any case, $b(\tau,P) \ge b(\kappa,P_1) + 1\ge\frac{1}{2}|\tau|$ 
and in turn $\tau$ is a good key by Lemma~\ref{lem:ub0}. 

\medskip
{\em Case 2:} The set of leaf descendants of $\ell_F(X_1)$ in $F - C(P_1)$ is some 
$Y \subseteq X_1 \setminus X^F_1(v'_0)$. In this case, the last cherry for $P$ to pick 
is $(Y, \{x_2\})$ and $P$ isolates $Y$ or $\{x_2\}$. By Lemma~\ref{lem:consistent}, 
$\ell_F(Y)$ is not a root in $F \ominus C(P_1)$. Thus, as in Case~1, 
we can prove that $\tau$ is a good key. 
\end{proof}

By Lemma~\ref{lem:dangle}, we may hereafter assume that for each $i\in\{1,2\}$, $\ell_F(X_i)$ 
is $X_{3-i}$-inclusive in $F$. By this assumption, $|X_1| \ge 2$ and $|X_2| \ge 2$. 
%
A vertex $v$ in $F$ is a {\em juncture} if one child $v_1$ of $v$ in $F$ is both 
$X_1$-inclusive and $X_2$-inclusive in $F$ and the other $v_2$ is $X_1$- or $X_2$-inclusive 
in $F$. A juncture $v$ in $F$ is {\em extreme} if no proper descendant of $v$ in $F$ is 
a juncture. Note that $\ell_F(X)$ is a juncture in $F$. So, extreme junctures exist in $F$. 

\begin{lemma}\label{lem:goodCut}
$(T,F)$ has a fair normal key $\kappa = (X_1, B)$ such that 
	$N_{B,X_2}(v) \le 1$ for every descendant $v$ of $\ell_F(X_2)$ in $F$.  
\end{lemma}
\begin{proof}
Roughly speaking, the proof is a refinement of that of Lemma~\ref{lem:goUp}. 
Let $I$ be the set of $X_1$-bifurcate vertices in $F$. The idea is to process the vertices 
of $I\cup X_1$ in a bottom-up fashion. We will maintain an invariant that once a $v\in I\cup X_1$ 
has been processed, we have obtained a fair normal key $\kappa_v = (X_1^F(v), B_v)$ of $(T,F)$ 
satisfying the following conditions:
\begin{itemize}
\item[I1.] $\kappa_v$ is robust if $N_{X_1}(v) = 0$. 
\item[I2.] $N_{B_v,X_2}(v) = 0$ if $v$ has a labeled descendant in $F-B_v$. 
\item[I3.] $N_{B_v,X_2}(u) \le 1$ for every descendant $u$ of $v$ in $F$. 
\end{itemize}

We first process each $x \in X_1$ by constructing $\kappa_x = (\{x\}, e_F(x))$. 
We then process those vertices $v \in I$ such that $v$ is $X_2$-exclusive in $F$ but 
its $X_1$-parent in $F$ is $X_2$-inclusive in $F$, by constructing $\kappa_v$ 
as in the proof of Lemma~\ref{lem:useful}; note that $\kappa_v$ is robust. 
Clearly, the invariant is maintained in the two cases. 
We next describe how to process each $X_2$-inclusive $v \in I$. Let $v_{1}$ and 
$v_{2}$ be the $X_1$-children of $v$ in $F$. For each $i\in\{1,2\}$, 
let $u_i$ be the child of $v$ in $F$ that is an ancestor of $v_i$ in $F$. 
We set $\kappa_v = (X_1^F(v), B_{v_1} \cup B_{v_2} \cup \{\tilde{e}\})$, 
where $\tilde{e}$ is chosen (in the listed order) as follows:
\begin{itemize}
\item[C1.] 
If $N_{B_{v_i},X_2}(u_i) \ge 1$ for some $i\in\{1,2\}$, then $\tilde{e} = e_F(u_i)$. 
({\em Comment:} By I2 in the invariant for $v_i$, each $X_2$-path in $F-B_{v_i}$ 
starting at $u_i$ must end at a leaf descendant of the head of some edge in $D_F(v,v_i)$. So, 
the nonexistence of semi-close stoppers for $(T,F)$ guarantees that $N_{B_{v_i},X_2}(u_i) = 1$ 
and $N_{B_{v_{j}},X_2}(u_{j}) = 0$, where $j=3-i$. Thus, I2 in the invariant for $v$ is 
maintained. Moreover, $\kappa_{v}$ is not robust only when $D_F(v,v_j) = \emptyset$ and 
$\kappa_{v_j}$ is not robust. Hence, by I1~in the invariant for $v_j$, $\kappa_v$ is 
not robust only when $N_{X_1}(v_j) \ge 1$ and in turn only when $N_{X_1}(v) \ge 1$. 
So, I1 in the invariant for $v$ is maintained. 
Because I2 in the invariant hold for both $v_1$ and $v_2$ and no descendant of $\ell_T(X_2)$ 
in $T$ is a semi-close stopper for $(T,F)$, I3 in the invariant for $v$ is maintained. 
By Lemma~\ref{lem:helpNormal}, $\kappa_v$ is a fair key.) 
\item[C2.] 
If for some $i\in\{1,2\}$, $D_F(v,v_i)=\emptyset$ and $\kappa_{v_i}$ is not robust, 
then $\tilde{e} = e_F(u_{3-i})$. 
\item[C3.] $\tilde{e} = e_F(u_i)$ for an arbitrary $i \in \{1,2\}$. 
\end{itemize}
Similar to and actually simpler than the above comment on C1, we can also comment on C2 and C3. 
So, the invariant for $v$ is maintained. We now define $\kappa = (X_1,B) = \kappa_{r}$, where 
$r = \ell_F(X_1)$. By the invariant, $\kappa$ is as desired. 
\end{proof}

\begin{lemma}\label{lem:noPure}
There is a good cut for $(T,F)$. 
\end{lemma}
\begin{proof}
Let $v$ be an extreme juncture in $F$, and $v_1$ and $v_2$ be the children of $v$ in $F$. 
Then, for some $i\in\{1,2\}$, both $v_1$ and $v_2$ are $X_i$-inclusive. 
We may assume $i=1$. 

For each $j\in\{1,2\}$, let $x_j$ be an $X_1$-descendant of $v_j$ in $F$. Let $\ddot{X}_2$ be 
the set of all $x \in X_2$ such that the head of some edge in $D_F(x_1,x_2)$ is an ancestor 
of $x$ in $F$. Since $v$ is $X_2$-inclusive in $F$, $\ddot{X}_2 \ne \emptyset$. Moreover, 
since $v$ is extreme, no edge in $D_F(x_1,x_2)$ is both $X_1$- and $X_2$-inclusive in $F$. 
So, $e_F(x) \in D_F(x_1,x_2)$ for each $x \in \ddot{X}_2$, because otherwise the head of 
some edge in $D_F(x_1,x_2)$ would be an $X_2$-port in $F$. Furthermore, since $v$ is extreme, 
there is at most one $x \in \ddot{X}_2$ with $e_F(x) \in D_F(x_j,v)$ for each $j\in\{1,2\}$. 
Thus, $1\le |\ddot{X}_2| \le 2$. 

First, consider the case where $|\ddot{X}_2| = 2$. Because $v$ is extreme, 
$e_F(x_3) \in D_F(x_1,v)$ for exactly one $x_3 \in \ddot{X}_2$, and $e_F(x_4) \in D_F(x_2,v)$ 
for exactly one $x_4\in \ddot{X}_2$. For each $j\in\{1,2\}$, $x_j$ and $x_{j+2}$ are siblings 
in $F$ because otherwise the sibling of $x_{j+2}$ in $F$ would be an $X_1$-port. 
So, by Lemma~\ref{lem:2cross}, $(T,F)$ has a good normal key and we are done. 

We hereafter assume that $\ddot{X}_2$ consists of a single vertex $x_3 \in X_2$. Then, 
there is an $x_4\in X_2$ that is not a descendant of $v$ in $F$ for $|X_2| \ge 2$. So, 
$e_F(x_j) \in D_F(x_3,x_4)$ for each $j\in\{1,2\}$, because otherwise the highest 
$X_2$-exclusive ancestor of $x_j$ in $F$ would be an $X_1$-port in $F$. Thus, for some 
$h\in\{1,2\}$, $x_h$ and $x_3$ are siblings in $F$ and $x_{3-h}$ is a child of $v$ 
in $F$. We may assume that $x_h = x_1$. Let $w$ be the parent of $x_1$ and $x_3$ in $F$. 
Since $v$ is extreme, each edge in $D_F(w,v)$ is $X$-exclusive. 

We construct a fair normal key $\kappa=(X_1,B)$ of $(T,F)$ as in the proof of 
Lemma~\ref{lem:goodCut}. We are {\em lucky} if at least one of the following holds:
\begin{itemize}
\item[D1.] For each component $K$ in $F\ominus B$, either the root of $K$ is a leaf or 
	both children of the root of $K$ are $X_2$-inclusive in $F\ominus B$. 
\item[D2.] $F\ominus B$ has at least two components $K$ such that 
	exactly one child of the root of $K$ is $X_2$-inclusive in $F\ominus B$. 
\end{itemize}

Suppose that we are lucky. Then, by Lemma~\ref{lem:2dis}, we can construct a good normal key 
$\kappa' = (X_2, B')$ of $(T,F\ominus B)$. Let $F' = F\ominus (B\cup B'')$, where $B''$ consists 
of the edges of $F$ corresponding to the edges in $B'$. 
Clearly, $|F'| = |F| + |B| + |B'|$ and $d(T,F') \le d(T,F) - \frac{|B|+|B'|-1}{2}$. 
We can modify $F'$ by adding back $w$ together with $e_F(x_1)$ and $e_F(x_3)$, i.e., by modifying 
$B\cup B''$ by replacing $e_F(x_1)$ and $e_F(x_3)$ with $e_F(w)$. The modification decreases $|F'|$ 
by~1 but does not increase $d(T,F')$. So, the modified $\hat{B}=B\cup B''$ is a cut of $F$ such that 
$d(T,F\ominus \hat{B}) \le d(T,F) - \frac{1}{2}|\hat{B}|$.

Next, suppose that we are unlucky. Then, $F\ominus B$ has exactly one component $K_1$ such that 
exactly one child of the root of $K_1$ is $X_2$-inclusive in $F\ominus B$. 
We want to modify $B$ so that we become lucky without violating the following important condition: 
$(X_1,B)$ is a fair normal key of $(T,F)$ and $N_{B,X_2}(u) \le 1$ for all vertices $u$ of $F-B$. 
In each modification of $B$ below, this important condition will be kept because of the nonexistence 
of semi-close stoppers for $(T,F)$; so, we will not explicitly mention this fact below. 

Recall that $x_2$ is a child of $v$ in $F$. So, by C2 in the proof of Lemma~\ref{lem:goodCut}, 
$e_F(v_1)$ is added to $B$ when we process $v$. 
Thus, $F\ominus B$ contains a component $K_2$ rooted at $v_1$. 
Since $(T,F)$ has no semi-close stopper, $|D_F(w,v)| \ge 2$ and 
in turn one child of $v_1$ in $K_2$ is $X$-exclusive. Hence, $K_1=K_2$. We have two ways to 
modify $B$. One is to replace $e_F(v_1)$ by $e_F(x_3)$ and the other is to replace $e_F(v_1)$ 
by the edge entering the child of $v_1$ that is an ancestor of $x_1$ in $F$. 
If modifying $B$ in the first way does not create a new component $K_3$ in $F\ominus B$ such that 
exactly one child of the root of $K_3$ is $X_2$-inclusive in $F\ominus B$, 
then we have become lucky because now $F\ominus B$ satisfies Condition~D1; otherwise, we become 
lucky by modifying $B$ in the second way because now $F\ominus B$ satisfies Condition~D2. 
\end{proof}

\subsection{Summarizing the Algorithm}
We here summarize our algorithm for finding a good cut for $(T,F)$. 
\begin{enumerate}
\item If there is a sibling-leaf pair $(x_1,x_2)$ in $T$ with $|D_F(x_1,x_2)|=1$, 
	then return $D_F(x_1,x_2)$. 

\item If there is a sibling-leaf pair $(x_1,x_2)$ in $F$ such that $|D_T(x_1,x_2)|=1$ 
	and the head of the unique edge in $D_T(x_1,x_2)$ is a leaf $x_3$, then return 
	$\{ e_F(x_3) \}$. 

\item If there is a root or disconnected stopper $\alpha$ for $(T,F)$, then compute 
	a good normal key $\kappa=(L_\alpha, B)$ for $(T,F)$ as in Lemma~\ref{lem:rootDis}, 
	and return $B$. 

\item If $(T,F)$ has an overlapping stopper $\alpha$ with children $\beta_1$ and $\beta_2$ 
	in $T$ such that there is an $L_{\beta_i}$-port in $F$ for some $i\in\{1,2\}$, then 
	compute a good normal key $\kappa=(L_\alpha, B)$ for $(T,F)$ as in Lemma~\ref{lem:port}, 
	and return $B$. 

\item If $(T,F)$ has an overlapping stopper $\alpha$ with children $\beta_1$ and $\beta_2$ 
	in $T$ such that $\ell_F(L_{\beta_i})$ is a descendant of $\ell_F(L_{\beta_{3-i}})$ 
	in $F$ and is $L_{\beta_{3-i}}$-exclusive in $F$ for some $i\in\{1,2\}$, then 
	compute a good cut $C$ for $(T,F)$ as in Lemma~\ref{lem:dangle}, and return $C$. 

\item If there is an overlapping stopper $\alpha$ for $(T,F)$, then compute 
	a good cut $C$ for $(T,F)$ as in Lemma~\ref{lem:noPure}, and return $C$.

\item Find a semi-close stopper $\alpha$ for $(T,F)$, use $\alpha$ to 
	compute a good abnormal key $\kappa=(L_\alpha, B, R)$ for $(T,F)$ as in 
	Lemma~\ref{lem:stopGoingUp}, and return $B$.
\end{enumerate}

\section*{Acknowledgments}
%
Lusheng Wang was supported by a National Science Foundation of China 
(NSFC 61373048) and a grant from the Research Grants Council 
of the Hong Kong Special Administrative Region, China [Project No. CityU 123013].


\begin{thebibliography}{99}

\bibitem{Bar+05}
Baroni, M., Grunewald, S., Moulton, V., and Semple, C. (2005)
Bounding the number of hybridisation events for a consistent evolutionary history. 
{\em Journal of Mathematical Biology}, {\bf 51}, 171-182.

\bibitem{BH06}
Beiko, R.G. and Hamilton, N. (2006)
Phylogenetic identification of lateral genetic transfer events, 
{\em BMC Evol. Biol.}, {\bf 6}, 159-169.

\bibitem{B+2006}
Bonet, M.L., John, K. St., Mahindru, R.,  and Amenta, N. (2006) 
Approximating subtree distances between phylogenies, 
{\em Journal of Computational Biology}, {\bf 13}, 1419-1434.

\bibitem{BMS08}
Bordewich, M., McCartin, C., and Semple, C. (2008) 
A 3-approximation algorithm for the subtree distance between
phylogenies. {\em Journal of Discrete Algorithms}, {\bf 6}, 458-471.

\bibitem{BS05} 
Bordewich, M. and Semple, C.  (2005)
On the computational complexity of the rooted subtree prune and regraft distance, 
{\em Annals of Combinatorics}, {\bf 8}, 409-423.

\bibitem{CW12}
Chen, Z.-Z., and Wang, L. (2012) 
FastHN: a fast tool for minimum hybridization networks, 
{\em BMC Bioinformatics}, 13:155. 

\bibitem{CW13}
Chen, Z.-Z., and Wang, L. (2013) 
An ultrafast tool for minimum reticulate networks. 
{\em Journal of Computational Biology}, 20(1): 38-41.

\bibitem{CFW15}
Chen, Z.-Z., Fan, Y., Wang, L. (2015)
Faster exact computation of rSPR distance. 
{\em Journal of Combinatorial Optimization}, 29(3): 605-635.

\bibitem{CMW16}
Chen, Z.-Z., Machida, E., Wang, L. (2016)
An Approximation Algorithm for rSPR Distance. 
{\em Proceedings of 22nd International Computing and Combinatorics Conference 
(COCOON'2016)}, Lecture Notes in Computer Science, Vol. 9797, pp. 468--479, 2016.

\bibitem{Shi+14}
Shi, F., Feng Q., You, J., Wang, J. (2014)
Improved Approximation Algorithm for Maximum Agreement Forest
of Two Rooted Binary Phylogenetic Trees. 
To appear in {\em Journal of Combinatorial Optimization}.




\bibitem{h+1996}
Hein, J., Jing, T., Wang, L., and Zhang, K.  (1996)
On the complexity of comparing evolutionary trees. 
{\em Discrete Appl. Math.}, {\bf 71}, 153-169.


\bibitem{MWZ99}
Ma, B., Wang, L., and Zhang, L. (1999)
Fitting distances by tree metrics with increment error. 
{\em Journal of Combinatorial Optimization}, {\bf 3}, 213-225.

\bibitem{MZ11}
Ma, B. and Zhang, L. (2011)
Efficient estimation of the accuracy of the maximum likelihood method for ancestral state 
reconstruction. {\em Journal of Combinatorial Optimization}, {\bf 21}, 409-422.


\bibitem{Mad97}
Maddison, W.P. (1997) Gene trees in species trees. 
{\em Systematic Biology}, {\bf 46}, 523-536.

\bibitem{Nak+05}
Nakhleh, L., Warnow, T., Lindner, C.R., and John, L.St. (2005)
Reconstructing reticulate evolution in species -- theory and practice. 
{\em Journal of Computational Biology}, {\bf 12}, 796-811.

\bibitem{RSW07}
Rodrigues, E.M, Sagot, M.-F., and Wakabayashi, Y. (2007)
The maximum agreement forest problem: Approximation algorithms and 
computational experiments. {\em Theoretical Computer Science}, {\bf 374}, 91-110.

\bibitem{Svv16}
Schalekamp, F., van Zuylen, A., and van der Ster, S. (2016)
A Duality Based 2-Approximation Algorithm for Maximum Agreement Forest. 
{\em Proceedings of ICALP 2016}, 70:1-70:14.



\bibitem{Wu09} Wu, Y.  (2009) 
A practical method for exact computation of subtree prune and regraft distance.
{\em Bioinformatics}, {\bf 25}(2), 190-196.

\bibitem{WBZ10} 
Whidden, C., Beiko, R. G., and Zeh, N.  (2010)
Fast FPT algorithms for computing rooted agreement forest: 
theory and experiments, {\em LNCS}, {\bf 6049}, 141-153.

\bibitem{WZ09}
Whidden, C. and Zeh, N. (2009) 
A unifying view on approximation and FPT of agreement forests. 
{\em LNCS}, {\bf 5724}, 390-401.
\end{thebibliography}
\end{document}